\documentclass[superscriptaddress,aps,prd,twocolumn,showpacs,nofootinbib]{revtex4-1}
\usepackage{graphicx}
\usepackage{amsmath,amssymb,bm}
\usepackage{color}

\begin{document}

\title{Superradiance in the sky}

\author{Jo\~ao G. Rosa \footnote{Also at Departamento de F\'{\i}sica e Astronomia, Faculdade de Ci\^encias da Universidade do Porto, Rua do Campo Alegre 687, 4169-007 Porto, Portugal.}}
\email{Email: joao.rosa@ua.pt\\ Also at Departamento de F\'{\i}sica e Astronomia, Faculdade de Ci\^encias da Universidade do Porto, Rua do Campo Alegre 687, 4169-007 Porto, Portugal.}
\affiliation{Departamento de F\'{\i}sica da Universidade de Aveiro and CIDMA, Campus de Santiago, 3810-183 Aveiro, Portugal} 

\begin{abstract}

We discuss the conditions under which plane electromagnetic and gravitational waves can be amplified by a rotating black hole due to superradiant scattering. We show, in particular, that amplification can occur for low-frequency waves with an incidence angle parametrically close to $0$ (or $\pi$) with respect to the black hole spin axis and with a parametrically small left (or right) polarization. This is the case of the radiation emitted by a spinning electric/magnetic dipole or gravitational quadrupole orbiting a black hole companion at large radius and co-rotating with the latter. This can yield observable effects of superradiance, for example, in neutron star-stellar mass black hole binaries, as well as in triple systems composed by a compact binary orbiting a central supermassive black hole. Due to superradiance, the total source luminosity in these systems exhibits a characteristic orbital modulation that may lead to significant observational signatures, thus paving the way for testing, in the near future, one of the most peculiar predictions of general relativity.
\end{abstract}


\maketitle


\section{Introduction}

Superradiance is one of the most interesting phenomena in wave scattering in black hole space-times. In general relativity, it is predicted to occur for bosonic waves (scalar, electromagnetic and gravitational) propagating in rotating and/or charged black holes\footnote{In this case only for charged bosonic fields.}, the former being the most interesting case from the astrophysical point of view and the focus of this work, since electromagnetic interactions between a black hole and the surrounding plasma should efficiently neutralize it.

Superradiant scattering is often referred to as the wave analogue of the Penrose process for energy extraction from a rotating black hole \cite{Penrose:1969pc, Christodoulou:1970wf}, although the nature of these processes is somewhat different. In the latter, a particle traveling in the ergoregion of a Kerr black hole may, for example, decay into two particles. Since in the ergoregion the Killing vector associated with time translations at infinity becomes space-like, one of the decay products may carry a negative energy into the black hole, decreasing its mass as measured by an asymptotic observer. The energy extracted must, by energy conservation, be carried away by the other decay product. In superradiant scattering (see \cite{review} for a recent review of this topic), a low frequency wave is amplified upon scattering off a rotating black hole, carrying away part of its energy and also angular momentum. This amplification occurs only for wave modes of the form $\psi\sim e^{-i\omega t+ im\phi}$, where $\phi$ is the azimuthal angle in Boyer-Lindquist coordinates, such that:
\begin{equation} \label{superradiance_condition}
\omega < m\Omega_H~,
\end{equation}
where $\Omega_H$ is the angular velocity of the black hole horizon. As we will discuss in more detail later on, this condition holds for the case where the wave frequency $\omega$ is positive-defined, such that the azimuthal wave number must be a positive integer, while there is an analogous condition for $\omega<0$ (i.e. modes of the form $e^{i\omega t}$) and $m<0$.

This phenomenon has been thoroughly analyzed in the literature following the seminal works of Zeldovich \cite{Zeldovich}, Starobinsky \cite{Starobinsky}, Teukolsky and Press \cite{Teukolsky:1973, Press:1973zz, Teukolsky:1974yv}, amongst others, in the 1970s, and has merited a significant attention in the recent literature since the realization that it could lead to instabilities at the linear level for Kerr black holes. This was first noted by Press and Teukolsky \cite{Press:1972zz} (and later studied in detail in \cite{Cardoso:2004nk}), who showed that by surrounding a rotating black hole with a reflective mirror a huge amount of energy could be extracted from the black hole due to multiple superradiant scatterings and reflections, a phenomenon popularly known as the ``black hole bomb". Amongst other mechanisms, this mirror-like effect occurs naturally for massive fields, which may be confined in the black hole's vicinity by the gravitational potential in quasi-bound states \cite{Damour:1976, Zouros:1979iw, Detweiler:1980uk, Furuhashi:2004jk, Cardoso:2005vk, Dolan:2007mj, Arvanitaki-JMR, Rosa:2009ei, Arvanitaki:2010sy, Rosa:2011my, Pani:2012bp, Rosa:2012uz, Dolan:2012yt, Brito:2013wya}. The superradiant condition (\ref{superradiance_condition}) requires the mass of the fields that can form these states around astrophysical black holes (stellar or supermassive) to be extremely small, namely below $10^{-10}$ eV, such that the black hole bomb mechanism could be a signature of exotic beyond the Standard Model particles such as axions \cite{Arvanitaki-JMR, Arvanitaki:2010sy}, hidden photons \cite{Rosa:2011my, Pani:2012bp} or massive gravitons \cite{Brito:2013wya}. It has also recently been shown that for frequencies at the superradiant threshold, $\omega=m\Omega_H$, the stable bound states found at the linear level have non-linear `hairy' black hole counterparts \cite{Herdeiro:2014goa}.

Despite this interesting connection to novel physics, with a potentially large phenomenological impact, superradiant scattering is in itself a remarkable prediction of general relativity in the Kerr geometry\footnote{Note that superradiance may occur in non-relativistic systems as the original Zeldovich cylinder \cite{Zeldovich}, and also possibly for rotating relativistic systems without event horizons \cite{Richartz:2013hza, Cardoso:2015zqa}.}. An observational confirmation of this process would therefore constitute an important test of Einstein's theory in a regime where it has yet to be probed. In \cite{Rosa:2015hoa}, we noted that pulsars constitute sources of low-frequency electromagnetic and gravitational radiation with the necessary properties to undergo superradiant scattering off a rotating black hole companion. We showed, in particular, that superradiance results in a modulation of the pulsar's luminosity in both electromagnetic and gravitational channels that could potentially be measured observationally.

In this work, besides giving more details on the process of superradiant scattering in pulsar-black hole binaries, we develop a more general analysis. In particular, we derive the generic conditions under which plane gravitational and electromagnetic waves can be amplified upon scattering off a rotating black hole. This is a non-trivial problem, since plane waves are superpositions of different modes of the form given above, some of which are amplified while others are attenuated upon scattering off a Kerr black hole. An overall amplification of a plane wave can therefore only occur when superradiant modes satisfying (\ref{superradiance_condition}) or its analogue for negative $\omega$ carry a sufficiently large fraction of the incident energy flux.

Plane waves are the leading form of the radiation emitted by a distant source, namely at a distance from the black hole exceeding the wavelength, and in this work we will also determine the properties that sources must exhibit such that their total luminosity increases upon scattering off a black hole companion. We then give examples of realistic sources that meet these requirements, in particular pulsars, as proposed in \cite{Rosa:2015hoa}, and also compact binaries. We note that our results provide only a first step towards testing the phenomenon of black hole superradiance, since we will not only focus on the plane-wave limit but also consider only the effects of superradiance on the total power of the source, rather than its luminosity along the line-of-sight or, in the case of gravitational waves, the associated strain. This involves additional technical difficulties that we hope to overcome in the future, but this work nevertheless lays the ground for studying more realistic astrophysical settings.

We will see that superradiant scattering of electromagnetic and gravitational radiation can be analyzed in a similar fashion and for completeness we discuss both cases in parallel. As already briefly discussed in \cite{Rosa:2015hoa}, the electromagnetic channel is not very promising from the observational point of view, since only very low-frequency radio waves can undergo superradiant scattering off astrophysical black holes. The required frequencies are, in particular, too low to be observed on Earth or its vicinity and may additionally not be able to propagate in the interstellar or intergalactic medium, so that one can hardly hope to observe any electromagnetic signature of superradiance. Gravitational radiation provides, on the other hand, a potentially very ``clean" channel, since it does not significantly interact with any astrophysical plasmas. The recent discovery of gravitational waves emitted by a coalescing binary black hole system \cite{Abbott:2016blz} provides, of course, a strong motivation for studying the effects of superradiance in this channel. We will show, in fact, that a compact binary is an ideal source of gravitational waves that can undergo superradiant scattering of a central supermassive black hole, yielding a very promising triple system (supermassive black hole - compact binary) for testing the superradiance phenomenon.
 
This work is organized as follows. In the next section, we give an overview of the phenomenon of black hole superradiance in the Kerr space-time and compute, in particular, the gain-loss factor upon scattering for different wave modes, reproducing the results of \cite{Teukolsky:1974yv}. In section III, we discuss the mode decomposition of plane waves and derive the general conditions under which the total energy flux can be amplified upon scattering off a Kerr black hole. We then apply these results to particular astrophysical sources satisfying the determined requirements, in particular pulsars and compact binaries, in section IV. We summarize our main conclusions and discuss possible extensions of this work in section V. In addition, we include two appendices on spin-weighted spherical harmonics and wave multipole decompositions.

In this work we use the $(+,-,-,-)$ metric convention and, by default, consider units such that $\hbar=c=G=1$ unless explicitly stated.  
 

\section{Superradiant scattering in the Kerr space-time}

In this section we review the basic concepts and techniques used to study the scattering of electromagnetic and gravitational radiation off a rotating Kerr black hole. We will consider a linear analysis, i.e.~neglecting the backreaction of the waves on the background space-time. This is suitable for most astrophysical sources where the energy extracted from the black hole, even over a very long period, is negligible compared to the black hole mass, as we will discuss later on when describing particular examples. We refer the reader to \cite{East:2013mfa} for a recent study of superradiant scattering at the non-linear level.

We start by writing the Kerr metric in Boyer-Lindquist coordinates \cite{Boyer:1966qh} for a black hole with ADM mass $M$ and spin $J=aM$:
\begin{eqnarray} \label{string_metric}
ds^2&=&\left(1-{2Mr\over \rho^2}\right)dt^2-{\rho^2\over\Delta}dr^2-\rho^2 d\theta^2\nonumber\\
&-&\left(r^2+a^2+{2Ma^2r\sin^2\theta\over \rho^2}\right)\sin^2\theta d\phi^2\nonumber\\
&+&{4Mr\over\rho^2}a\sin^2\theta dtd\phi~,
\end{eqnarray}
where $\Delta=r^2+a^2-2Mr$ and $\rho^2=r^2+a^2\cos^2\theta$. In these coordinates, the event horizon is located at $r_+=M+\sqrt{M^2-a^2}$ and the inner Cauchy horizon at $r_-=M-\sqrt{M^2-a^2}$, such that $\Delta(r_\pm)=0$. The ergoregion is bounded by the surface $r_e=M+\sqrt{M^2-a^2\cos^2\theta}$ and corresponds to the region where the Killing vector $\xi_t=\partial/\partial t$, which is associated with time-translations at infinity, becomes space-like and negative energy states may exist from the perspective of an asymptotic observer.

The study of wave propagation in this background geometry is greatly simplified by using the Newman-Penrose (NP) formalism \cite{Newman:1961qr}, where one projects the electromagnetic field tensor and the Weyl tensor on a null tetrad, typically chosen to be the Kinnersley tetrad \cite{Kinnersley:1969zza}  defined by the 4-vectors $(l^\mu, n^\mu, m^\mu, \bar{m}^\mu)$:
\begin{eqnarray} \label{Kinnersley_tetrad}
l^\mu&=&\left[{r^2+a^2\over \Delta},1,0,{a\over\Delta}\right]~,\nonumber\\
n^\mu&=&{1\over2\rho^2}\left[r^2+a^2,-\Delta,0,a\right]~,\nonumber\\
m^\mu&=&{1\over\sqrt{2}\bar{\rho}}\left[ia\sin\theta,0,1,i/\sin\theta\right]~,
\end{eqnarray}
where $\bar\rho=r+ia\cos\theta$. Note that $l^\mu$ and $n^\mu$ coincide with the two principal null directions of the Kerr metric's Weyl tensor, which makes this tetrad particularly suitable for the study of incoming and outgoing radiation. For electromagnetic waves, we can project the Maxwell tensor onto this tetrad to obtain the complex NP scalars:
\begin{eqnarray} \label{NP_scalars_EM}
\phi_0=F_{\mu\nu}l^\mu m^\nu~, \qquad \phi_2=F_{\mu\nu}m^{*\mu} n^\nu~.
\end{eqnarray}
Analogously, for gravitational radiation we project the Weyl tensor on the tetrad to obtain the complex scalars:
\begin{eqnarray} \label{NP_scalars_grav}
\psi_0&=&-C_{\mu\nu\rho\sigma}l^\mu m^\nu l^\rho m^\sigma~, \nonumber\\
 \psi_4&=&-C_{\mu\nu\rho\sigma}n^\mu \bar{m}^\nu n^\rho \bar{m}^\sigma~.
\end{eqnarray}
Note that other scalar quantities can be obtained for both types of waves, but these will not be relevant for our discussion. It was shown by Teukolsky \cite{Teukolsky:1973} that both Maxwell's equations and the perturbed Einstein equations in the Kerr space-time lead to decoupled second order differential equations for each of the NP scalars and that, in particular, the equations for $\phi_0$, $\phi_2$, $\psi_0$ and $\psi_4$ can be solved by separation of variables. More explicitly, we can perform a mode decomposition of the form:
\begin{eqnarray} \label{NP_mode_expansion}
\Upsilon_s&=&\sum_{l,m,\omega} e^{-i\omega t+im\phi}{}_sS_{slm}(\theta){}_sR_{lm}(r)~,
\end{eqnarray}
where $s=1$ corresponds to $\phi_0$, $s=-1$ to $\bar\rho^2\phi_2$, $s=+2$ to $\psi_0$ and $s=-2$ to $\bar{\rho}^4\psi_4$. The NP scalars thus yield the $s=\pm1$ and $s=\pm2$ components of electromagnetic and gravitational waves, respectively. The corresponding angular functions satisfy the equation for spin-$s$ spheroidal harmonics, which up to $\mathcal{O}(a\omega)$ terms coincide with the corresponding spin-weighted spherical harmonics, i.e.
\begin{equation} \label{spheroidal_harmonics}
 e^{im\phi}{}_sS_{lm}(\theta)= {}_sY_{lm}(\theta,\phi)+ \mathcal{O}(a\omega)
 \end{equation}
 with eigenvalues $\lambda=l(l+1)-s(s+1)+\mathcal{O}(a\omega)$. The radial equation for each spin-$s$ wave mode is then given by Teukolsky's master equation:
\begin{eqnarray} \label{radial_equation}
& &\Delta{d^2{}_sR_{lm}\over dr^2}+2(s+1)(r-M){d{}_sR_{lm}\over dr} \nonumber\\
&+&\left({K^2-2is(r-M)K\over\Delta}+4is\omega r-\lambda\right){}_sR_{lm}=0~,
\end{eqnarray}
where $K(r)=(r^2+a^2)\omega-ma$.


\subsection{Analytical asymptotic matching}

Teukolsky's equation cannot be solved exactly with known analytical methods, but several approximate procedures have been developed. A particularly useful technique is to match the exact solutions that one can obtain in two overlapping regions: (i) the near region $r-r_+\ll \omega^{-1}$ and (ii) the far region $r-r_+\gg r_+$. An overlap between these two regions, and hence a consistent matching, is then possible for $\omega r_+\ll 1$ \cite{Starobinsky}.

In the near-region, the general solution is given in terms of hypergeometric functions, and upon imposing ingoing boundary conditions at the horizon, such that no waves escape from within the black hole, we obtain, omitting the $(s,lm)$ indices for simplicity:
\begin{eqnarray} \label{near_solution}
R_{near}&=& A_{hole}(x+\tau)^{i\varpi/\tau}x^{-s-i\varpi/\tau}\nonumber\\
&\times &{}_2F_1(-l,l+1,1-s-2i\varpi/\tau,-x/\tau)~,
\end{eqnarray}
where we have defined the normalized distance to the horizon $x=(r-r_+)/r_+$, the extremality parameter $\tau=(r_+-r_-)/r_+$ ($0\leq\tau\leq1$) and $\varpi=(2-\tau)(\bar\omega-m\bar\Omega_H)$, with barred quantities being normalized to the horizon radius, e.g.~$\bar\omega=\omega r_+$. This implies that close to the horizon we have $R\sim \Delta^{-s}e^{-i\omega r_*}$ in terms of the tortoise coordinate defined via $dr_*/dr= (r^2+a^2)/\Delta$.

In the far-region, the solution is given in terms of confluent hypergeometric functions:
\begin{eqnarray} \label{far_solution}
R_{far}&=& e^{-i\bar\omega x}x^{l-s}\left[C M(l+1-s, 2l+2, 2i\bar\omega x)\right.\nonumber\\
&+&\left. D x^{-2l-1}M(-l-s,-2l,2i\bar\omega x)\right]~.
\end{eqnarray}
This yields an expansion in terms of incoming and outgoing waves at infinity:
\begin{eqnarray} \label{asymptotic_solution}
R_{far}(r)\rightarrow  A_{in}{e^{-i\omega r}\over r}+ A_{out}{e^{i\omega r}\over r^{2s+1}}~,
\end{eqnarray}
where 
\begin{eqnarray} \label{asymptotic_coeff}
A_{in}&=&\left[C(-2i\bar\omega)^{-l+s-1}{\Gamma(2l+2)\over\Gamma(l+1+s)}\right.\nonumber\\
&+&\left.D(-2i\bar\omega)^{l+s}{\Gamma(-2l)\over \Gamma(-l+s)}\right]r_+~,\nonumber\\
A_{out}&=&\left[C(2i\bar\omega)^{-l-1-s}{\Gamma(2l+2)\over\Gamma(l-s+1)}\right.\nonumber\\
&+&\left.D(2i\bar\omega)^{l-s}{\Gamma(-2l)\over \Gamma(-l-s)}\right]r_+^{2s+1}~.
\end{eqnarray}
Taking the limits $x\gg 1, \tau$ of the near-region solution and $\bar\omega x\ll 1$ of the far-region solution, we obtain that both solutions are given in terms of two monomials in $x^{l-s}$ and $x^{-l-s-1}$. Matching the coefficients of each monomial in the near- and far-region solutions yields:
\begin{eqnarray} \label{matching_condition}
{D\over C}={\Gamma(-2l-1)\over\Gamma(-l)}{\Gamma(l+1)\over\Gamma(2l+1)}{\Gamma(l+1-s-2i\varpi/\tau)\over\Gamma(-l-s-2i\varpi/\tau)}\tau^{2l+1}~,\nonumber\\
\end{eqnarray}
which we may use to determine the ratio between the incoming and outgoing energy far away from the black hole. For electromagnetic waves, we can see from Eq.~(\ref{asymptotic_solution}) that $\phi_0$ ($s=1$) gives to leading order the incoming wave, while $\phi_2$ ($s=-1$) yields the outgoing wave. Similarly, in the gravitational case $\psi_0$ ($s=+2$) is an incoming wave and $\psi_4$ ($s=-2$) is an outgoing wave.  Writing the electromagnetic energy-momentum tensor in terms of the NP scalars \cite{Chandrasekhar}, one can show that at infinity:
\begin{eqnarray} \label{incoming_outgoing_energy_EM}
{d^2E_{in}\over dt d\Omega}=\lim_{r\rightarrow +\infty} r^2{|\phi_0|^2\over 8\pi}~, \ \ \ \ {d^2E_{out}\over dt d\Omega}=\lim_{r\rightarrow +\infty} r^2{|\phi_2|^2\over 2\pi}~,\nonumber\\
\end{eqnarray}
where the difference in numerical factors is due to the additional $1/2$ normalization of $n^\mu$ compared to $l^\mu$. Similarly, for gravitational waves one can define a pseudo-energy-momentum tensor that yields:
\begin{eqnarray} \label{incoming_outgoing_energy_GW}
{d^2E_{in}\over dt d\Omega}=\lim_{r\rightarrow +\infty} r^2{|\psi_0|^2\over 64\pi\omega^2}~, \ \ \ {d^2E_{out}\over dt d\Omega}=\lim_{r\rightarrow +\infty} r^2{|\psi_4|^2\over 4\pi\omega^2}~.\nonumber\\\!\!\!
\end{eqnarray}
Note that asymptotically the space-time is flat, such that in terms of the electromagnetic fields $\phi_0=(\mathbf{E}+i\mathbf{B})\cdot\mathbf{e}_+$ and $2\phi_2=(\mathbf{E}+i\mathbf{B})\cdot\mathbf{e}_-$, where $\mathbf{e}_\pm= (\mathbf{e}_{\hat\theta}\pm i\mathbf{e}_{\hat\phi})/\sqrt{2}$ in the orthonormal spherical basis, which justifies the above expressions. Similarly, for waves propagating along the radial direction $\psi_0=-(\omega^2/8)h_{ij}\mathbf{e}_+^{ \hphantom{+} i}\mathbf{e}_+^{ \hphantom{+} j}$ and $\psi_4=-(\omega^2/2)h_{ij}\mathbf{e}_-^{\hphantom{-} i}\mathbf{e}_-^{\hphantom{-} j}$ in terms of the metric perturbations $h_{ij}$ in the transverse-traceless gauge. 

The two NP scalars are, in each case, related by linear differential equations that take a simple form at infinity \cite{Press:1973zz}. We may then express the incoming and outgoing power in terms of a single quantity, which we choose as $\phi_2$ for the electromagnetic case and $\psi_4$ for the gravitational case. Integrating over the solid angle, we then have for the incoming and outgoing power at infinity for electromagnetic waves:
\begin{eqnarray} \label{incoming_outgoing_energy_EM}
{dE_{in}\over dt }={16\omega^4\over l^2(l+1)^2}\left|A_{in}\right|^2~, \qquad
{dE_{out}\over dt }=|A_{out}|^2\,,
\end{eqnarray}
while in the gravitational case:
\begin{eqnarray} \label{incoming_outgoing_energy_GW}
{dE_{in}\over dt }&=&{128\omega^6\over l^2(l+1)^2(l-1)^2(l+2)^2}\left|A_{in}\right|^2~, \nonumber\\
{dE_{out}\over dt }&=&{1\over 2\omega^2}|A_{out}|^2~.
\end{eqnarray}
With these results, we may use the relations (\ref{asymptotic_coeff}) and the matching condition (\ref{matching_condition}) to obtain the overall gain/loss factor, which we can write for a general spin $s$ wave mode with angular numbers $(l,m)$ in the form:
\begin{eqnarray} \label{fractional_gain}
{}_sZ_{lm}(\omega)&\equiv& {dE_{out}/dt \over dE_{in}/dt}-1\nonumber\\
&\simeq& -2(\bar\omega-m\bar\Omega_H){(2-\tau)\over\tau}(2\bar\omega\tau)^{2l+1}\nonumber\\ 
&\times &\left[{(l+s)!(l-s)!\over (2l)!(2l+1)!}\right]^2\prod_{k=1}^l \left(k^2+{4\varpi^2\over \tau^2}\right)\,,
\end{eqnarray}
which is analogous to the expression first obtained in \cite{Starobinsky}. In obtaining the last line of the previous equation we have used that $\bar\omega\tau\ll 1$ and that:
\begin{eqnarray} \label{aux_gamma_results}
{\Gamma(-n)\over\Gamma(-m)}&=&(-1)^{n-m}{m!\over n!}~, \nonumber\\
 {\Gamma(l+2+x)\over \Gamma(-l+2+x)}&=&-{l+1\over l}(-1)^l x\prod_{k=1}^l(k^2-x^2)~.
\end{eqnarray}
We note that an analogous expression can be obtained by computing the energy flux into the black hole horizon, and by conservation of energy we have ${}_sZ_{lm}=-(dE_{hole}/dt)/ (dE_{in}/dt)$. In addition, we must take into account that both the electromagnetic field and the metric are real quantities,  although the NP scalars are complex since the components of the tetrad 4-vectors $m^\mu$ and $\bar{m}^\mu$ are complex-valued. This implies that realistic waves will include modes of the form $e^{-i\omega t}$ and $e^{i\omega t}$, which can be implemented by considering both positive and negative values for $\omega$, with the absolute frequency of the wave given by $|\omega|$ in this notation. It is easy to repeat the analysis above for $\omega<0$ to conclude that: 
\begin{eqnarray} \label{fractional_gain_negative}
{}_sZ_{lm}(-\omega)={}_sZ_{l,-m}(\omega)~.
\end{eqnarray}
Superradiance is explicit in the form of the gain/loss factor ${}_sZ_{lm}(\omega)\propto -(\omega- m\Omega_H)$, such that the reflected (outgoing) power exceeds the incoming power for $\omega <m\Omega_H$. For $\omega>0$, superradiant amplification, ${}_sZ_{lm}>0$, will then occur for $m>0$ modes satisfying the condition (\ref{superradiance_condition}), while for $\omega<0$ only the $m<0$ modes will be amplified.

 The result above shows that for low (positive) frequencies $\omega \ll r_+^{-1}$ superradiance will be more effective for the lowest (co-rotating) multipoles, namely $l=m=1$ and $l=m=2$ in the electromagnetic and gravitational cases, respectively, where to leading order we obtain:
\begin{eqnarray} \label{gain_loss_factor}
{}_1Z_{1,1}&\simeq& -{4\over 9}(\bar\omega-\bar\Omega_H)\bar\omega^3 (2-\tau)\tau^2~,\nonumber\\
{}_2Z_{2,2}&\simeq& -{4\over 225}(\bar\omega-2\bar\Omega_H)\bar\omega^5 (2-\tau)\tau^4~.
\end{eqnarray}
In this limit, superradiance is thus more effective in extracting energy and spin from the black hole for electromagnetic waves. The result above also suggests that ${}_sZ_{lm}=0$ for extremal black holes, $\tau=0$, but this case has to be treated separately since the near-region solution (\ref{near_solution}) is not well-defined in this limit. The matching procedure is, moreover, limited to low-frequency waves and numerical solutions of the Teukolsky equation show that superradiant amplification is much stronger close to extremality and frequencies just below the superradiant threshold, yielding a maximum amplification of 4.4\% for electromagnetic waves ($\omega\simeq 0.88\Omega_H$) and 138\% for gravitational waves ($\omega\simeq 2\Omega_H$) in the corresponding lowest multipoles \cite{Teukolsky:1974yv}. In the next subsection we then develop a numerical method to compute the gain/loss factor for different wave modes.


\subsection{Numerical analysis}

Teukolsky's equation can be solved numerically by using a forward integration method \cite{Press:1973zz}, that can be simply implemented with Mathematica. This method consists in starting with an ingoing wave arbitrarily close to the horizon of the black hole and then integrating Teukolsky's equation numerically up to a large distance, where the coefficients of the incoming and outgoing waves can be extracted for different frequencies.

The ingoing boundary condition at the horizon can be set by noting that the radial function ${}_sR_{lm}$ admits a generic power-series expansion of the form:
\begin{eqnarray} \label{horizon_expansion}
{}_sR_{lm}= x^{-s-i\varpi/\tau}\sum_{n=0}^{\infty}a_n x^n~.
\end{eqnarray}
The series coefficients $a_n$ can be easily determined by substituting this series expansion into the radial equation and solving iteratively the resulting algebraic equations. We can then integrate the Teukolsky equation up to a distance $x\gg \bar\omega^{-1}$, where as seen earlier the solution has the asymptotic form:
\begin{eqnarray} \label{asymptotic_solution_x}
{}_sR_{lm}(x)\rightarrow  {{}_sA_{in}^{lm}\over r_+}{e^{-i\bar\omega x}\over x}+ {{}_sA_{out}^{lm}\over r_+^{2s+1}}{e^{i\bar\omega x}\over x^{2s+1}}~.
\end{eqnarray}
From this it is clear that the incoming waves give the leading behaviour for $s=+1$ and $s=+2$ ($\phi_0$ and $\psi_0$ respectively), so that it will be more convenient to consider these NP scalars in the numerical procedure. The coefficient ${}_sA_{in}^{lm}$ can then be extracted by evaluating $x|{}_sR_{lm}(x)|$ at a sufficiently large distance.

We can then compute the energy flow into the black hole and compare it with the incoming power. We have for the energy flow into the black hole \cite{Press:1973zz}, setting the overall normalization $a_0=1$ for both electromagnetic and gravitational waves:
\begin{eqnarray} \label{energy_hole_num}
{dE_{hole}\over dt}=
\begin{cases}
{\tau^2\bar\omega r_+^2\over 4\varpi}~, & s=+1~,\\
{\tau^4\bar\omega r_+^4\over 32\varpi (\varpi^2+\tau^2/4)}~, &  s=+2
\end{cases}~.
\end{eqnarray}
For the incoming energy, we obtain:
\begin{eqnarray} \label{incoming_hole_num}
{dE_{in}\over dt}=
\begin{cases}
{r_+^2\over 4}|{}_{+1}A_{in}^{lm}|^2~, & s=+1~,\\
{r_+^4\over 32\bar\omega^2}|{}_{+2}A_{in}^{lm}|^2~, &  s=+2
\end{cases}~.
\end{eqnarray}
The gain/loss factor defined in Eq.~(\ref{fractional_gain}) is thus given by:
\begin{eqnarray} \label{fractional_gain_num}
{}_sZ_{lm}=
\begin{cases}
-{\tau^2\bar\omega\over\varpi}|{}_{+1}A_{in}^{lm}|^{-2}~, & s=+1~,\\
-{\tau^4\bar\omega^3\over \varpi(\varpi^2+\tau^2/4)}|{}_{+2}A_{in}^{lm}|^{-2}~, &  s=+2
\end{cases}~.
\end{eqnarray}
For numerical accuracy, it is important to include spheroidal corrections to the angular eigenvalues:
\begin{eqnarray} \label{angular_eigenvalues_num}
\lambda=l(l+1)-s(s+1)-2a m\omega+a^2\omega^2+\sum_{k=1}^{+\infty}c_k(a\omega)^k~,\nonumber\\
\end{eqnarray}
where approximate analytical expressions for the $c_k$ coefficients can be found in \cite{Berti:2005gp}, with numerical values tabulated in \cite{Press:1973zz, Teukolsky:1974yv}. We note that in \cite{review} a fully numerical analysis in both the radial and angular directions was implemented.

\begin{figure}[htbp]\vspace{-0.4cm}
\centering\includegraphics[scale=1]{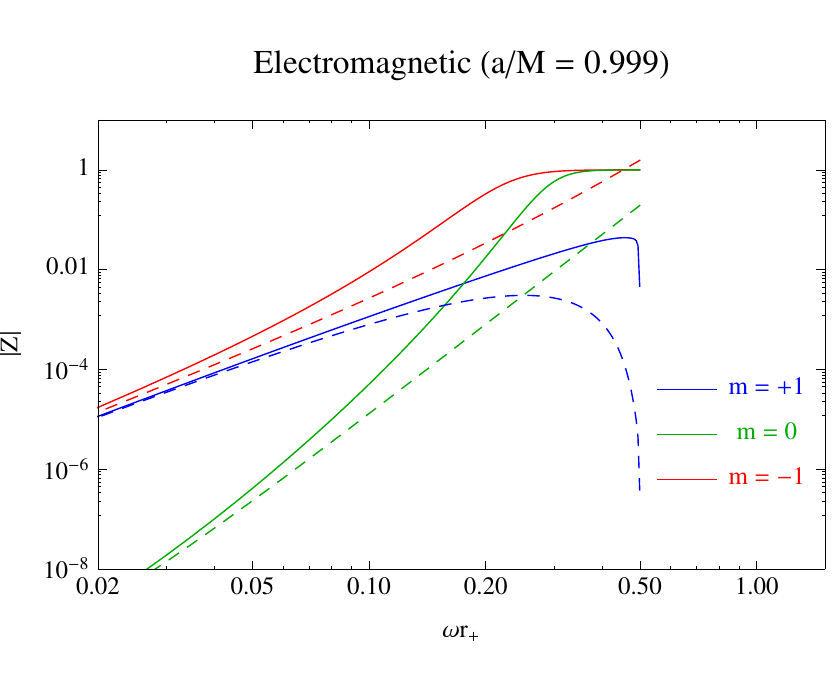}
\centering\includegraphics[scale=1]{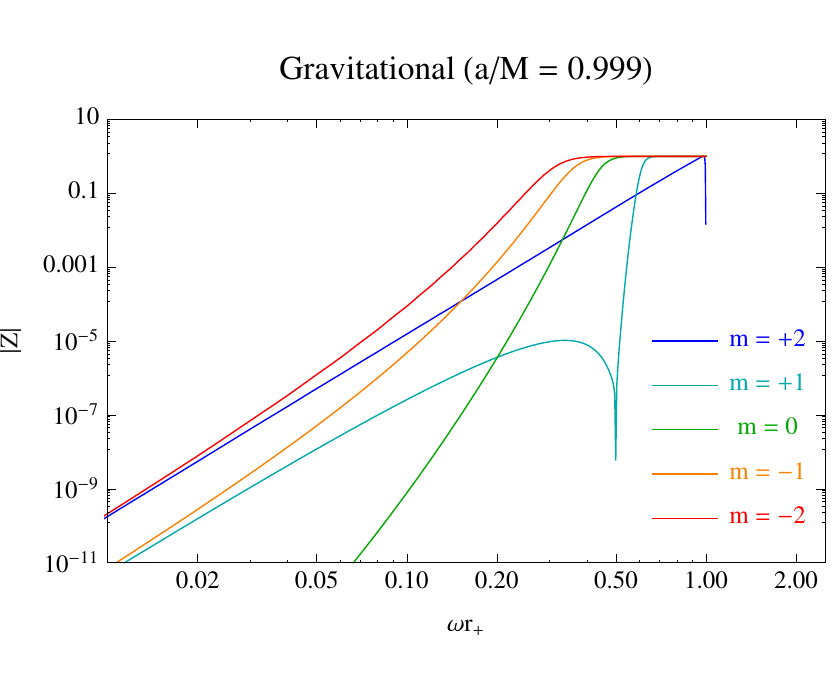}
\caption{Numerical results for the gain/loss factor for a near-extremal black hole with $a/M=0.999$, for the lowest multipoles $l=1$ and $l=2$ in the electromagnetic and gravitational cases, respectively. Note that $Z<0$ for $m\leq0$ and also when $\omega>m\Omega_H$ for $m>0$ (with $\omega>0$). In the electromagnetic case we also show the approximate asymptotic matching prediction for low frequency (dashed lines).}
\label{numerical}
\end{figure}

 In Figure \ref{numerical} we show the results obtained for $a/M=0.999$ for the leading multipoles of electromagnetic and gravitational waves.

As one can see in this figure, the gain/loss factor is well described by the analytical expressions obtained earlier using the asymptotic matching procedure for small frequencies. These are given by the dashed curves in the electromagnetic case and this is also observed for gravitational waves, although for clarity of the plot we do not show these explicitly. 

As the frequency increases, the analytical expressions generically underestimate $|{}_sZ_{lm}|$ for the co-rotating modes, while for the non-superradiant modes $m\leq 0$ this only occurs up to a given frequency, since the gain/loss factor saturates at high frequencies to $Z\simeq-1$. This implies that the incident energy is almost completely absorbed by the black hole, in agreement with the results obtained by Press and Teukolsky \cite{Teukolsky:1974yv}, and is related to the fact that such waves have a sufficiently high energy to overcome the angular momentum barrier that suppresses absorption/amplification for lower frequencies.

Co-rotating modes are, as expected for $\omega>0$, amplified in the superradiant regime $\omega <m\Omega_H$ and the amplification factor increases with the black hole spin and frequency until very close to the superradiant threshold, particularly for gravitational waves. As opposed to the behavior observed at low frequencies, gravity waves are amplified by a much larger factor than electromagnetic waves close to the superradiant threshold, as well as for larger spins close to extremality. For $a/M=0.999$, we obtain a maximum amplification factor $Z_{EM}^{max}=0.436$ for electromagnetic waves and $Z_{GW}^{max}=1.02$ for gravitational waves. Press and Teukolsky obtained a maximum of 1.38 for gravitational wave scattering off a black hole with spin $a/M=0.99999$, and in this case we obtain 1.33, in fairly good agreement  with \cite{Teukolsky:1974yv}\footnote{We note that our numerical code requires the inclusion of a large number of terms in the near-horizon series Eq.~(\ref{horizon_expansion}) very close to extremality, being hard to evaluate the accuracy of the result for $a/M=0.99999$. It would be useful to use a different numerical procedure to better understand the small discrepancy of our results with those of  \cite{Teukolsky:1974yv} in this regime.}.

We emphasize, in particular, that for electromagnetic waves the maximum amplification for $l=m=1$ is considerably smaller than the corresponding absorption coefficient  $|Z|\simeq 1$ for $m=0,-1$. For gravitational waves, on the other hand, superradiant amplification of the $l=m=2$ mode is of the same order (and may even be slightly larger) than the non-superradiant absorption of the $m\leq 0$ modes.


\section{Plane wave decomposition and conditions for amplification}

Having obtained the form of the gain-loss factor for different wave modes, we can now investigate under which conditions an incident wave will be amplified, which of course depends on the relative abundance of superradiant and non-superradiant modes in the incident flux. We will focus our discussion on incident plane waves, which constitute a good approximation for sources that are sufficiently far away from the black hole, namely when the distance $L$ between the source and its spinning black hole companion is large compared to the wavelength of the radiation, $\lambda$. This approximation also requires $L\gg r_+, R_S$, where $R_S$ is the source radius (or typical size if not spherical). Our scattering problem will then have an incident plane wave far away from the black hole, at distances $r_+ \ll r \ll L$, rather than explicitly considering a source placed at a finite distance from the horizon in the Kerr space-time.

The problem we pose is then the following. Given a plane wave boundary condition at infinity (i.e. $r\gg r_+$), with a given incident direction $(\theta_0,\phi_0)$ in Boyer-Lindquist coordinates, what is the fraction of the incident energy flux carried by superradiant and non-superradiant modes? Knowing the answer to this question and that different $(\omega, l , m)$ modes evolve independently and are amplified/attenuated by the gain/loss factors computed in the previous section, we can compute the energy output of the scattering process. This will allow us to determine what are the properties of the incident plane wave required for an overall amplification and energy extraction from the black hole. 

We should note at this stage that in realistic problems the source will not be static and, in particular, we will be interested in cases where it orbits the black hole companion. Since orbital frequencies, $\Omega_{orb}$, are typically much smaller than the source and black hole angular frequencies, denoted by $\omega_S$ and $\Omega_H$, respectively, we may use an adiabatic approximation, i.e.~treat the source as fixed in computing the outcome of a scattering process. 

Since the boundary conditions for the scattering problem are posed far away from the black hole, we may use a flat space approximation to determine the multipolar decomposition of an incident plane wave. Let us first consider the simpler electromagnetic case, where the electric and magnetic fields for a generic plane wave of frequency $\omega$ can be written as:
\begin{eqnarray} \label{plane_em}
\mathbf{E}&=&{1\over 2}e^{-i\omega t+ i\mathbf{k}\cdot \mathbf{r}} \left[\epsilon_1\mathbf{e}^{(1)}+\epsilon_2\mathbf{e}^{(2)}~\right]+ \mathrm{c.c.}~,\nonumber\\
 \mathbf{B}&=&\mathbf{n}\times \mathbf{E}~,
\end{eqnarray}
where $\mathbf{n}=\left(\sin\theta_0\cos\phi_0,\sin\theta_0\sin\phi_0,\cos\theta_0\right)^t$ denotes the unit 3-vector in the direction of propagation and $\mathbf{k}=\omega \mathbf{n}$. We can obtain the polarization vectors by starting with a plane wave traveling along the $z$-direction and performing two rotations, first about the $x$-axis by an angle $\theta_0$ and then by an angle $\phi_0-\pi/2$ about the $z$-axis. The unit vectors along the $x$- and $y$-axes then become:
\begin{eqnarray} \label{em_polarizations}
\mathbf{e}^{(1)}&=&\left(\sin\phi_0,-\cos\phi_0,0\right)^t~,\nonumber\\
\mathbf{e}^{(2)}&=&\left(\cos\theta_0\cos\phi_0,\cos\theta_0\sin\phi_0,-\sin\theta_0\right)^t~.
\end{eqnarray}
Note that the electric and magnetic fields are real quantities, but the polarization amplitudes $\epsilon_1$ and $\epsilon_2$ are in general complex. One can also see in Eq.~(\ref{plane_em}) that because of the reality condition a plane electromagnetic wave will include modes with time-dependence $e^{-i\omega t}$ and $e^{i\omega t}$, i.e.~positive and negative frequencies according to the definition given above. As we have seen in the previous section, these behave differently in a scattering process in the Kerr space-time, which is why we have separated their contributions in the form of the electric and magnetic fields. This is also true for gravitational waves as we discuss below.

As we have seen above, to study the scattering problem it is convenient to use the NP formalism and compute the NP scalars associated with the electromagnetic field above. It is easy to show, in particular, that
\begin{eqnarray} \label{phi2_plane}
\phi_2&=&-{2\pi i\over3}\left[\epsilon_Re^{-i\omega t+ i\mathbf{k}\cdot \mathbf{r}}+\epsilon_L^*e^{i\omega t- i\mathbf{k}\cdot \mathbf{r}}\right]\nonumber\\
&\!\! \times &\!\!\!\!\!\!  \sum_{m=-1}^{+1}\!\!  \!\! {}_{-1}Y_{1, m}^*(\theta_0,\phi_0){}_{-1}Y_{1, m}(\theta,\phi)~,
\end{eqnarray}
where $\epsilon_{L,R}=(\epsilon_1\pm i\epsilon_2)/\sqrt{2}$ are the circular polarization factors and ${}_s Y_{lm}(\theta,\phi)$ are spin-weighted spherical harmonics, in this case for $s=-1$. A similar expression can be obtained for $\phi_0$ with $s=+1$ spherical harmonics. We provide in appendix A a list of the lowest order spin-$s$ spherical harmonics for $s=\pm1 $ and $s=\pm 2$.

From Eq.~(\ref{phi2_plane}) it would seem that the wave corresponds to a superposition of only $l=1$ angular modes, but one must take into account the angular dependence included in the factor $e^{i\mathbf{k}\cdot \mathbf{r}}$, which has the well-known decomposition in terms of scalar spherical harmonics:
\begin{eqnarray} \label{scalar_harmonics}
e^{i\mathbf{k}\cdot \mathbf{r}} = 4\pi \sum_{l,m} i^l Y_{lm}^*(\theta_0,\phi_0)Y_{lm}(\theta,\phi) j_l(kr)~, 
\end{eqnarray}
where $j_l(x)$ are spherical Bessel functions of integer order and $k= |\mathbf{k}|=\omega$ in natural units. The NP scalar thus corresponds to a superposition of products of spin-1 and spin-0 spherical harmonics, and one can use the Clebsch-Gordan coefficients given in appendix B to show that, at large distances, $kr\gg 1$, one has:
\begin{eqnarray} \label{phi2_large}
\phi_2\!\!&\simeq&\!\!-2\pi i \epsilon_Re^{-i\omega t} \sum_{lm}\left[a_{1l}^{(in)} {e^{-ikr}\over (kr)^3}+a_{1l}^{(out)}{e^{ikr}\over kr}\right]\nonumber\\
&\times & {}_{-1}Y_{lm}^*(\theta_0,\phi_0){}_{-1}Y_{lm}(\theta,\phi)\nonumber\\
& +& \left(\omega\leftrightarrow -\omega, \epsilon_R\leftrightarrow \epsilon_L^*\right)~,
\end{eqnarray}
where $a_{1l}^{(in)}= (-1)^l l(l+1)/8$ and $a_{1l}^{(out)}=-i/2$. Hence, at large distances from the black hole (compared to the wavelength of the radiation), the NP scalar is a superposition of $s=-1$ harmonics, each with an incoming and an outgoing component. The amplitude of each multipole is given by the corresponding harmonic in the direction of propagation, ${}_{-1}Y_{lm}^*(\theta_0,\phi_0)$. One also sees that positive and negative frequency modes have the same multipolar decomposition but with a generically different overall amplitude, proportional to the right and left circular polarization factors, respectively.

We can then use Eq.~(\ref{incoming_outgoing_energy_EM}) to compute the incoming part of the energy flux in the plane wave for each mode, which yields the boundary condition for the scattering problem. Integrating over the solid angle, we thus obtain the remarkably simple expression:
\begin{eqnarray} \label{incident_flux_EM}
{dE_{in}^{(l,m)}\over dt} &=& {\pi^2\over\omega^2}\left|\epsilon_{R,L}\right|^2\left|{}_{-1}Y_{lm}(\theta_0,\phi_0)\right|^2~.
\end{eqnarray}

We can proceed in an analogous fashion for plane gravitational waves, which in the transverse-traceless gauge can generically be written in the form:
\begin{eqnarray} \label{plane_gw}
h_{ij}={1\over 2}e^{-i\omega t+ i\mathbf{k}\cdot \mathbf{r}} \left[h_+\mathbf{e}_{ij}^++h_\times\mathbf{e}_{ij}^\times~\right]+ \mathrm{c.c.},
\end{eqnarray}
where $i,j$ denote spatial components. The `plus' and `cross' polarization tensors associated with a wave propagating in the direction of the unit vector $\mathbf{n}$ given above can be obtained by rotating the corresponding tensors for a wave propagating along the $z$ direction, as described above in the electromagnetic case, being given by:
%
\begin{eqnarray} \label{gw_polarizations}
\mathbf{e}_{ij}^{+}&=&\left(
\begin{tabular}{c c c}
$s^2_{\phi_0}-c^2_{\theta_0} c^2_{\phi_0}$ & $-(1+c^2_{\theta_0})c_{\phi_0} s_\phi$ & $s_{\theta_0} c_{\theta_0} c_{\phi_0}$\\
$-(1+c^2_{\theta_0})c_{\phi_0} s_{\phi_0}$ & $c^2_{\phi_0}-c^2_{\theta_0} s^2_{\phi_0}$ &  $s_{\theta_0} c_{\theta_0} s_{\phi_0}$\\
 $s_{\theta_0} c_{\theta_0} c_{\phi_0}$ &  $s_{\theta_0} c_{\theta_0} s_{\phi_0}$ & $-s^2_{\theta_0}$
\end{tabular}\right)~,\nonumber\\
\mathbf{e}_{ij}^{\times}&=&\left(
\begin{tabular}{c c c}
$c_{\theta_0} s_{2\phi_0}$ & $- c_{\theta_0} c_{2\phi_0}$ & $-s_{\theta_0} s_{\phi_0}$\\
$-c_{\theta_0} c_{2\phi_0}$ & $-c_{\theta_0} s_{2\phi_0}$ & $s_{\theta_0} c_{\phi_0}$\\
$-s_{\theta_0} s_{\phi_0}$ & $s_{\theta_0} c_{\phi_0}$  & 0
\end{tabular}\right)~,
\end{eqnarray}
%
where $c_\alpha \equiv \cos\alpha$ and $s_\alpha \equiv \sin\alpha$.  These tensors satisfy the transversality and orthogonality relations $e_{ij}^{+} n_j = e_{ij}^{\times} n_j= e_{ij}^{+}e_{ji}^{\times}=0$.

We may then compute the NP scalar $\psi_4$ for this plane wave in flat space, yielding:
\begin{eqnarray} \label{psi4_plane}
\psi_4&\!\!=&\!\!{\sqrt{2}\pi\over5}\omega^2\left[h_Re^{-i\omega t+ i\mathbf{k}\cdot \mathbf{r}}+h_L^*e^{i\omega t- i\mathbf{k}\cdot \mathbf{r}}\right]\nonumber\\
&\times &\sum_{m=-2}^{+2}{}_{-2}Y_{2, m}^*(\theta_0,\phi_0){}_{-2}Y_{2, m}(\theta,\phi)~.\nonumber\\
\end{eqnarray}
This is similar to what we obtained for the electromagnetic case but with $s=-2$ spherical harmonics, and with the circular polarization factors defined as $h_{L,R}=(h_+\pm ih_\times)/\sqrt{2}$. As above, using the scalar multipolar decomposition (\ref{scalar_harmonics}) and the Clebsh-Gordan coefficients given in appendix B, we obtain at large distances:
\begin{eqnarray} \label{psi4_large}
\psi_4&\simeq& \sqrt{2}\pi h_R\omega^2 e^{-i\omega t} \sum_{lm}\left[a_{2l}^{(in)} {e^{-ikr}\over (kr)^5}+a_{2l}^{(out)}{e^{ikr}\over kr}\right]\nonumber\\
&\times& {}_{-2}Y_{lm}^*(\theta_0,\phi_0){}_{-2}Y_{lm}(\theta,\phi)\nonumber\\
& +& \left(\omega\leftrightarrow -\omega, h_R\leftrightarrow h_L^*\right)~,
\end{eqnarray}
where $a_{2l}^{(in)}= (-1)^l(l-1)l(l+1)(l+2)/32$ and $a_{2l}^{(out)}= -i/2$. We thus find an analogous result for plane electromagnetic and gravitational waves, with the latter being given by a superposition of $s=-2$ spherical harmonics with amplitude modulated by the corresponding harmonic in the direction of propagation, each multipole with an incoming and an outgoing radial component. We can then use Eq.~(\ref{incoming_outgoing_energy_GW}) to compute the incident energy carried by each mode, yielding the simple formula:
\begin{eqnarray} \label{incident_flux_GW}
{dE_{in}^{(l,m)}\over dt} &=& {\pi^2\over 4}\left|h_{R,L}\right|^2\left|{}_{-2}Y_{lm}(\theta_0,\phi_0)\right|^2~,
\end{eqnarray}
with, as above, the right (left) polarization corresponding to positive (negative) frequency modes. We note that for both types of radiation the incoming and outgoing fluxes are equal, but the latter would correspond to an unscattered wave.

\begin{figure}[htbp]
\includegraphics[scale=1]{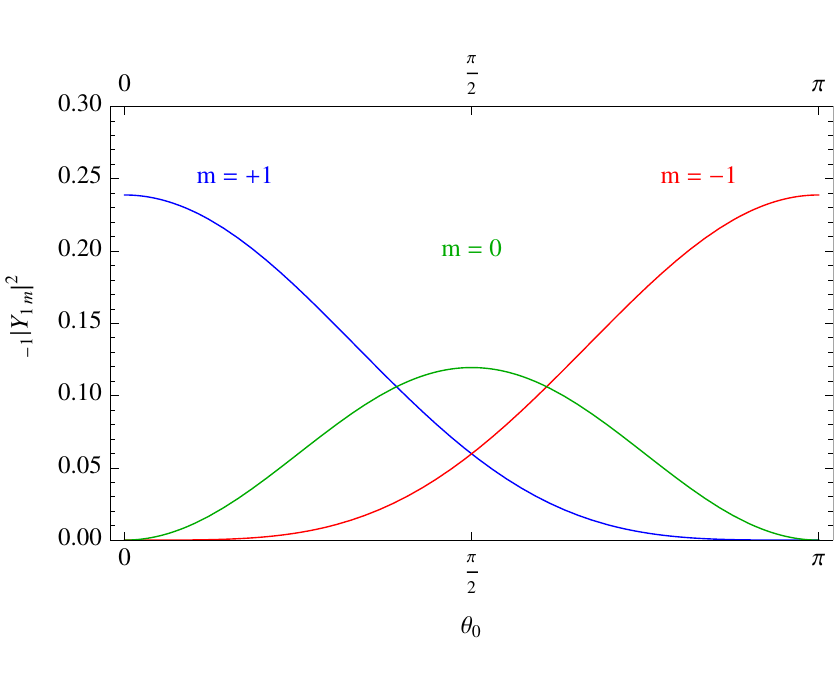}\vspace{-0.5cm}
\includegraphics[scale=1]{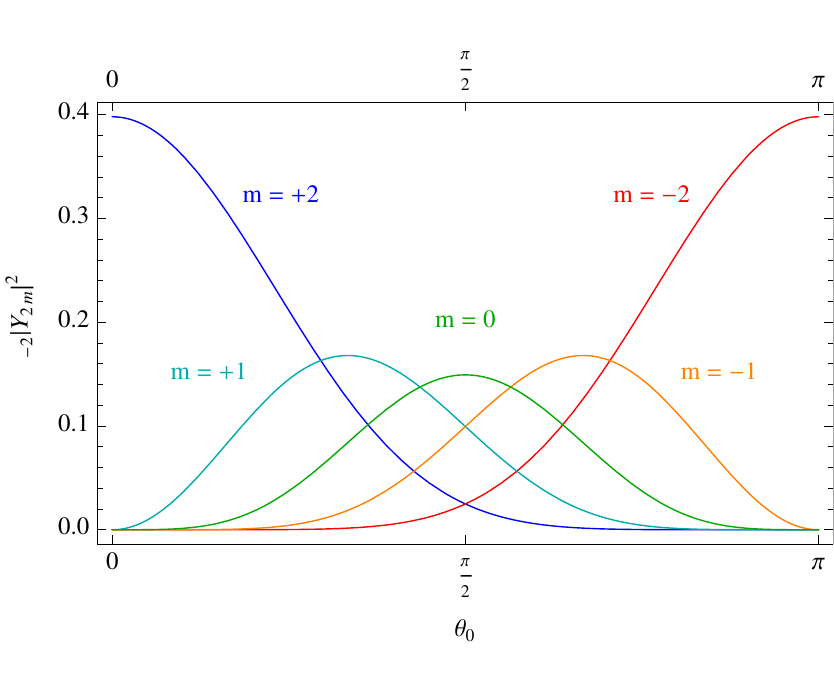}\vspace{-0.5cm}
\caption{The square modulus of the spin-weighted spherical harmonics  $\left|{}_{s}Y_{lm}(\theta_0,\phi_0)\right|^2$ for ($s=-1$, $l=1$) and ($s=-2$, $l=2$) as a function of the angular direction $\theta_0$ of an electromagnetic and gravitational plane wave, respectively.}
 \label{harmonics}
\end{figure}

We thus see that in both cases one obtains a simple formula for the incoming energy flux at infinity for each $(\omega, l ,m)$ multipole in a plane wave. This leads us to two important conclusions. Firstly, the energy flux for different multipoles is proportional to the square modulus of the corresponding spin-weighted spherical harmonic in the propagation direction, implying in particular that different $m$ modes will be dominant for different directions. We illustrate this in Fig.~\ref{harmonics}, where we plot the relevant functions for the lowest multipoles, $l=1$ and $l=2$, in the electromagnetic and gravitational cases, respectively.

The behavior illustrated in this figure is generic for higher multipoles, and we see that positive (negative) $m$ modes are dominant at small (large) angles, whereas close to the equatorial plane they carry a comparable amount of the incoming flux, with the $m=0$ modes dominating in this case. Hence, for positive (negative) frequencies, superradiant modes will dominate at small (large) angles.

Secondly, the relative abundance of positive and negative frequency modes in the plane wave may also be different, since independently of the multipole $(l,m)$ their amplitude is given by the generically distinct right and left polarizations, respectively. This is crucial for an overall amplification of the wave. Consider, for example, the case of linearly polarized waves where $|\epsilon_{R}|=|\epsilon_{L}|$ or $|h_{R}|=|h_{L}|$, such that both positive and negative frequency modes have the same amplitude. If e.g.~a wave propagates in the $\theta_0=0$ direction, it will be composed only of $m>0$ modes. The positive frequency modes will then be amplified, while the negative frequency modes will be attenuated. We have seen in the previous section that gain/loss factors for superradiant modes can be slightly above unity for the lowest multipoles in the gravitational case, but that for similar frequencies most non-superradiant modes are maximally attenuated. Hence, one cannot get an overall amplification of the wave in this case, as shown in \cite{Dolan:2008kf} for gravitational waves, and it is not difficult to convince oneself that the same reasoning applies for other incidence angles $\theta_0$.

To obtain an overall amplification of a plane wave, we thus need {\it chiral waves}, in the sense that the right and left polarization factors must be distinct. Since superradiant modes have either $\omega>0$ and $m>0$ or $\omega<0$ and $m<0$, we must furthermore require that the right polarization (positive frequency) dominates at small incidence angles, or that the left polarization (negative frequency) dominates at large angles. It is also clear that for incidence angles close to $\theta_0=\pi/2$ (i.e.~along the black hole's equatorial plane) it will be very hard to get an overall amplification, since positive and negative $m$ modes have similar amplitudes.

In summary, and given that superradiant amplification factors are larger for the lowest multipoles, an overall amplification of plane electromagnetic or gravitational waves requires the following conditions to be satisfied:

\begin{enumerate}
\item low-frequency waves with $\omega\lesssim \Omega_H$;

\item sufficiently large or small incidence angles $\theta_0$;

\item right/left circular polarizations dominating at  small/large incidence angles.
\end{enumerate}

Regarding the first condition, we note that the horizon's angular velocity for a Kerr black hole is given by:
\begin{eqnarray} \label{black_hole_rotation}
\Omega_H&=&{ac\over {r_+^2+a^2}}={c^3\over 2GM}\left({\tilde{a}\over 1+\sqrt{1-\tilde{a}^2}}\right)\\\nonumber
&\simeq& 102\ \mathrm{kHz}\left({M_\odot \over M}\right) \left({\tilde{a}\over 1+\sqrt{1-\tilde{a}^2}}\right)~,
\end{eqnarray}
where we have defined the dimensionless spin parameter $\tilde{a}= J c/ GM^2$. On the one hand, astrophysical black holes in the stellar mass range, i.e.~with $\mathcal{O}(10-100)$ solar masses, have angular velocities of at most a few kHz. On the other hand, supermassive black holes, which are thought to exist in the center of most active galaxies, have masses in the range $10^6M_\odot-10^9 M_\odot$ and hence angular velocities in the range $10^{-4}-10^{-1}$ Hz. Thus, any astrophysical source of electromagnetic or gravitational radiation that is amplified by a black hole companion must emit in these frequency ranges (for near-extremal black holes) or at lower frequencies. We thus need to look for very low-frequency sources of electromagnetic and gravitational waves satisfying conditions (2) and (3). 

The second condition is essentially a constraint on the type of orbit followed by the source around its black hole companion. On the one hand, if the orbital plane is not sufficiently inclined with respect to the black hole's equatorial plane, one cannot expect an overall amplification of the radiation to occur anywhere in the orbit. For orbits with a large inclination, on the other hand, amplification may occur at least in some points in the orbit where the incidence angle deviates sufficiently from $\pi/2$.

The third condition is perhaps the least obvious in an astrophysical context, since the wave polarization must be related to the incidence angle in a particular way. It nevertheless points towards spinning sources, which emit radiation that is not linearly polarized. We will see in the next section in analyzing particular sources that the simplest cases satisfying this condition are spinning magnetic/electric dipoles and mass quadrupoles whose spins are aligned with the black hole's rotation axis. 

We note that these conditions only apply to massless waves with a non-zero spin. The scattering of putative scalar waves can be analyzed in a similar fashion, and it is easy to conclude that, since the scalar spherical harmonics $Y_{l,\pm m}(\theta,\phi)$ have the same $\theta$ dependence, modes with opposite $m$ will always be on equal footing independently of the incidence angle. Superradiant modes can thus never dominate over non-superradiant modes in the incident flux, and a plane scalar wave will always be attenuated in the Kerr space-time, in agreement with \cite{Macedo:2013afa}.


\section{Astrophysical sources exhibiting superradiant amplification}

Having determined the generic conditions under which plane waves can be amplified by a spinning black hole, we will now discuss particular examples of astrophysical systems where superradiance may occur. We will consider two examples: a pulsar-black hole binary system and a triple system composed of a compact binary orbiting a central supermassive black hole. The first case was already discussed in \cite{Rosa:2015hoa} and is relevant for black holes in the stellar mass range. Here we will give more details of the analysis performed in \cite{Rosa:2015hoa} and put them in the context of the general framework discussed in the previous section. The second case is a new proposal relevant for supermassive black holes, which require sources of much lower frequency radiation than that emitted by known pulsars. We will see that the basic principles of the analysis are common to both examples, paving the way to find other potential examples of astrophysical sources that may exhibit superradiant behavior when orbiting a black hole companion. 

The generic configuration of the systems that we will analyze is shown in the figure below, where we give the coordinates of the source (taken as fixed in the adiabatic approximation discussed above) in the black hole's frame and its relevant geometrical properties.
\vspace{-0.5cm}

\begin{figure}[htbp]
\hspace{-3cm}\vspace{-1.5cm}\includegraphics[scale=0.4]{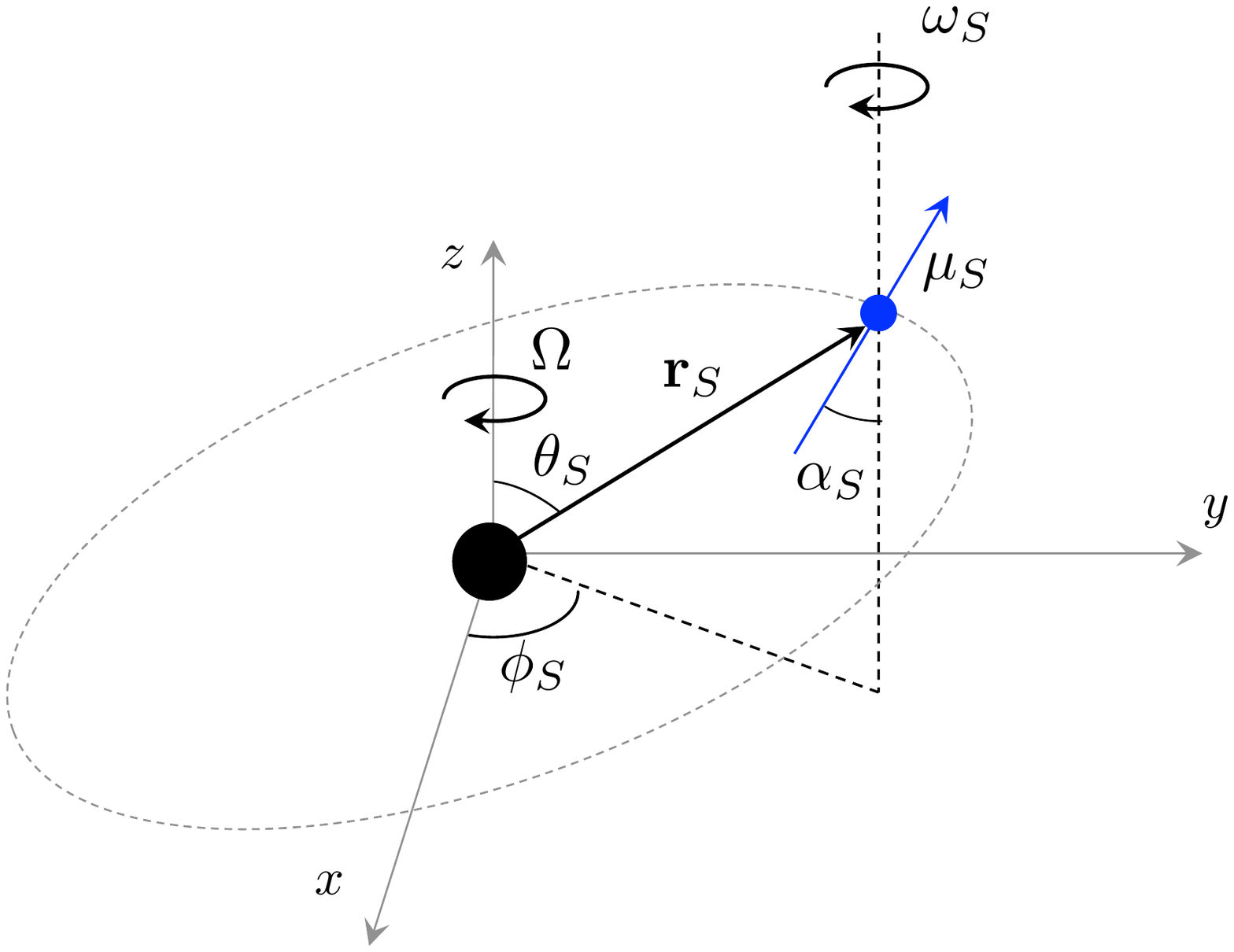}
\caption{Generic configuration of a source (blue circle) orbiting a central black hole (black circle at the origin) at position $\mathbf{r}_S= (L, \theta_S, \phi_S)$ in Boyer-Lindquist coordinates. The black hole and source are spinning with frequencies $\Omega_H$ and $\omega_S$, respectively, and their rotation axes are aligned with the $z$ direction (we will also discuss the anti-aligned configuration). The blue arrow indicates the relevant source moment (magnetic dipole, mass quadrupole), which generically makes an angle $\alpha_S$ with the spin axis.}
\end{figure}

Since we are interested in the case of large orbital distances, $L\gg \lambda, r_+, R_S$, we will take the source to be in flat space as the leading approximation, consistently with the analysis of the previous section. 


\subsection{Pulsar-black hole binaries}

Pulsars are spinning neutron stars that are believed to result from supernova explosions. They emit electromagnetic radiation, typically in the radio band, in a narrow beam along their magnetic axis, which does not typically coincide with the rotation axis, producing a ``lighthouse" effect such that a pulsar is observed as a periodic source. These periodic pulses can be timed with high-accuracy, making pulsars one of the most precise clocks known in the Universe and ideal for astrophysical tests of general relativity \cite{Kramer:2004gi}. Neutron stars are typically highly magnetized, as a consequence of magnetic flux conservation during the collapse of the parent star, and are believed to primarily lose (rotational) energy through the emission of magnetic dipole radiation at its rotation frequency. Mili-second pulsars would thus emit electromagnetic radiation in the right frequency range to undergo superradiant scattering off a stellar mass black hole companion, although as we will discuss later on this radiation may interact with surrounding plasmas, making it difficult for this process to occur in realistic systems. In fact, we cannot observe the magnetic dipole radiation emitted by a pulsar from the Earth, since its frequency is generically below the average plasma frequency in the atmosphere. The observable radio emission is actually the result of charge acceleration in the pulsar's magnetosphere.

However, neutron stars should also typically exhibit a non-vanishing mass quadrupole moment, due to small deviations from axial symmetry arising from a variety of possible sources that we discuss below. Hence, they also emit gravitational waves at (twice) its rotation frequency, again in the right range for superradiant scattering with stellar mass Kerr black holes. We discuss below the properties of both the electromagnetic and gravitational radiation emitted by a pulsar in the frame of the black hole companion and analyze the prospects for observing the effects of superradiance in realistic systems.


\subsubsection{Magnetic dipole radiation}

Although a neutron star's magnetosphere can be rather complex, its main properties are, to leading order, well approximated by a magnetic dipole precessing about its rotational axis, which is known as the {\it oblique-rotator} model \cite{Pacini, Poisson}.

Considering that the magnetic dipole moment of the neutron star $\mathbb{\mu}_S= \mathbf{m}_P$ makes an angle $\alpha$ with its rotation axis, which we assume to be aligned with the $\mathbf{\hat{z}}$-direction, and rotates about it with the pulsar's rotational frequency, $\omega_S=\omega_P$, we have:
\begin{eqnarray} \label{pulsar_moment}
\mathbf{m}_P=\left({m_0\over 2}\sin\alpha e^{-i\omega_S t} \left[\mathbf{\hat{x}}\pm i\mathbf{\hat{y}}\right]+ \mathrm{c.c.} \right)+m_0\cos\alpha \mathbf{\hat{z}}~,\nonumber\\
\end{eqnarray}
where $m_0$ denotes the absolute magnitude of the dipole and the upper (lower) sign corresponds to a pulsar co-(counter-)rotating with the black hole. Hence, for $\alpha\neq 0 $, the transverse components of the dipole will oscillate sinusoidally in time. Since the neutron star's radius is typically much smaller than the wavelength of the emitted radiation, i.e.~well within the near-zone for electromagnetic radiation emission, we can write the surface magnetic field as:
\begin{eqnarray} \label{pulsar_magnetic_field}
\mathbf{B}={\mu_0\over4\pi}{3(\mathbf{m}_P\cdot \mathbf{n'})\mathbf{n'}-\mathbf{m}_P\over R_P^3 }~,
\end{eqnarray}
where $R_P$ is the pulsar radius and $\mathbf{n'}$ denotes the unit vector along the radial direction from the centre of the star. The magnitude of the surface magnetic field thus reaches a maximum at the poles, where $\mathbf{n'}$ is aligned with the dipole moment, yielding:
\begin{eqnarray} \label{pulsar_magnetic_field_max}
B_{\mathrm{max}}={\mu_0\over 4\pi}{2m_0\over R_P^3}~.
\end{eqnarray}
Away from the pulsar, in the far region, the electric and magnetic fields have the standard dipole form:
\begin{eqnarray} \label{pulsar_radiation_far}
\mathbf{B}&=&-{\mu_0\over 4\pi}{e^{ik|\mathbf{r}-\mathbf{r}_S}|\over |\mathbf{r}-\mathbf{r}_S|}\left(\mathbf{n}\times \ddot{\mathbf{m}}_P\right)\times \mathbf{n}~,\nonumber\\
\mathbf{E}&=&{\mu_0 c\over 4\pi}{e^{ik|\mathbf{r}-\mathbf{r}_S}|\over |\mathbf{r}-\mathbf{r}_S|}\left(\mathbf{n}\times \ddot{\mathbf{m}}_P\right)~,
\end{eqnarray}
where $\mathbf{n}= (\mathbf{r}-\mathbf{r}_S)/ |\mathbf{r}-\mathbf{r}_S|$ is the distance from the source and $k=\omega_P/c$. Taking the limit where the pulsar is far away from the black hole, $L\gg |\mathbf{r}|$, we have:
\begin{eqnarray} \label{plane_wave_limit}
{e^{ik|\mathbf{r}-\mathbf{r}_S|}\over |\mathbf{r}-\mathbf{r}_S|}\simeq {e^{ikL}\over L}e^{i\mathbf{k}\cdot \mathbf{r}}~,
\end{eqnarray}
and the electromagnetic field can then be written in the plane wave form given in Eq.~(\ref{plane_em}) with frequency $\omega=\omega_S$, direction $-\hat{\mathbf{r}}_S$ and polarization factors:
\begin{eqnarray} \label{pol_pulsar_em}
\epsilon_1&=& {\mu_0 c m_0\sin \alpha\omega_S^2\over 4\pi L}e^{ikL} e^{i\phi_S} \cos\theta_S~, \nonumber\\
\epsilon_2&=& \mp i{\mu_0 c m_0\sin \alpha\omega_S^2\over 4\pi L}e^{ikL} e^{i\phi_S} ~, 
\end{eqnarray}
such that
\begin{eqnarray} \label{pol_pulsar_em_comb}
|\epsilon_R| &=& {\mu_0 c m_0\sin \alpha\omega_S^2\over \sqrt{6\pi } L}|{}_{-1}Y_{1, \pm1}(\theta_0, \phi_0)|~, \nonumber\\
|\epsilon_L| &=& {\mu_0 c m_0\sin \alpha\omega_S^2\over \sqrt{6\pi } L}|{}_{-1}Y_{1, \mp1}(\theta_0, \phi_0)|~, 
\end{eqnarray}
where we note that the wave's angular direction is $\theta_0=\pi-\theta_S$ and $\phi_0=\pi+\phi_S$, i.e.~opposite to the pulsar's position vector. We thus see that the magnetic dipole radiation emitted by the spinning neutron star can have a chiral polarization in the sense defined above, and that the polarization depends on the wave direction and the direction of the pulsar's spin. That the polarization factors $\epsilon_{R,L}$ are proportional to the spin-weighted spherical harmonics ${}_{-1}Y_{1,\pm1}(\theta_0,\phi_0)$ is a direct consequence of the dipolar nature of the source. Furthermore, when the pulsar co-rotates with the black hole, we see that the positive (negative) frequency modes in the plane wave will be dominant for small (large) angles, coinciding with the dominance of positive (negative) $m$ modes as we had seen earlier. Thus, when the pulsar's angular position deviates sufficiently from the equatorial plane, its radiation will be dominantly amplified. The opposite occurs when the pulsar and black hole spins are anti-aligned, and the dominant modes are never superradiant in this case.

To quantify this effect, we may first use the electromagnetic field in Eq.~(\ref{pulsar_radiation_far}) at large $r$ to compute the associated NP scalar $\phi_2$ and the expression for the outgoing energy flux in Eq.~(\ref{incoming_outgoing_energy_EM}) (including missing constants) to obtain the total pulsar's luminosity in the absence of the black hole, yielding the well-known magnetic dipole formula:
\begin{eqnarray} \label{pulsar_power_EM}
P_{EM}&=& {\omega_p^4 \mu_0\over 6\pi c^3}m_0^2\sin^2\alpha\nonumber\\
&\simeq&  2.5\times 10^9\!\left(1\ \mathrm{ms}\over T_P\right)^{4}\!\!\left({B_{\mathrm{max}}s_\alpha\over 10^8\ \mathrm{T}}\right)^2\!\!\left(R_P\over 10\ \mathrm{km}\right)^6 L_\odot~,\nonumber\\
\end{eqnarray}
where we have considered the typical properties of a mili-second pulsar, with $T_P$ denoting the rotational period and $L_\odot$ the solar luminosity, showing that a pulsar emits a large amount of energy in magnetic dipole radiation. 

Combining the results above with the expression for the incoming energy flux in each multipole obtained in Eq.~(\ref{incident_flux_EM}), we find:
\begin{eqnarray} \label{incoming_mode_energy_pulsar_EM}
{dE_{in}^{(l,m)}\over dt} =\pi{\lambda^2\over L^2}\left|{}_{-1}Y_{1,\pm1}(\theta_0,\phi_0)\right|^2\left|{}_{-1}Y_{lm}(\theta_0,\phi_0)\right|^2P_{EM}\nonumber\\
\end{eqnarray}
for the positive frequency modes and an analogous expression for the negative frequency modes with $\pm \rightarrow \mp$. This expression yields the boundary condition for each multipole in our scattering problem. We may then multiply this by the gain/loss factor in Eq.~(\ref{fractional_gain}) to obtain the outgoing flux upon scattering off the rotating black hole companion, or equivalently the energy flux through the black hole horizon. The pulsar's luminosity will then correspond to the one computed in the absence of the black hole subtracted of the energy flux through the black hole's horizon, which may yield an increase or a decrease in the pulsar's luminosity, depending on whether energy is extracted from the black hole or not, respectively.

\begin{figure}[htbp]
\includegraphics[scale=1]{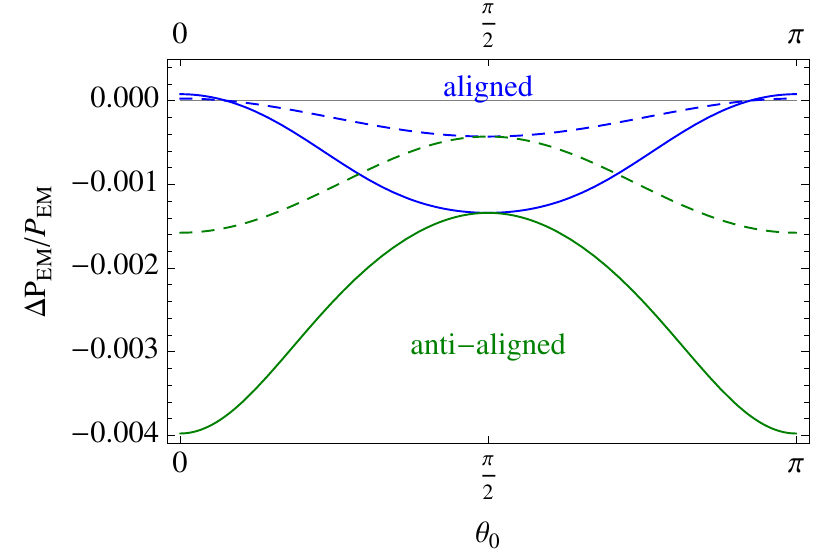}
\caption{Fractional variation of the pulsar's total magnetic dipole luminosity due to a near-extremal black hole companion with $\tilde{a}=0.999$, for a pulsar rotational frequency $\omega_S\simeq 0.94\Omega_H$ (solid curves) and for $\tilde{a}=0.9$, $\omega_S\simeq 0.87\Omega_H$ (dashed curves). The blue (green) curves correspond to a co-rotating (counter-rotating) binary system, and in all cases the orbital distance was taken to be $L=10\lambda$.}
\label{modulation_em}
\end{figure}

In Fig.~\ref{modulation_em} we show an example of this effect for a near-extremal black hole with $\tilde{a}=0.999$ and a pulsar source with $\omega_S\simeq 0.94\Omega_H$ for which one obtains the maximal amplification for the $l=m=1$ multipole, which is the most amplified mode. We take the pulsar to be at a distance $L=10\lambda$ from the black hole so as to be within the plane wave limit. We show the quantity $\Delta P_{EM}/P_{EM}$ corresponding to the fractional variation of the pulsar's luminosity due to the presence of the black hole, computed using the $l=1,2,3$ multipoles, with higher multipoles yielding a negligible contribution.

As one can observe in this figure, an overall amplification of the pulsar's luminosity can only occur in the case where the pulsar and the black hole rotate in the same direction and when the pulsar is sufficiently far away from the equatorial plane (at $\theta_S=\pi/2$), in particular with angular deviations from the equator exceeding $\sim 76.5^\circ$. Even at these angular positions, the amplification effect is very small, specifically about one part in $10^4$ in this example. The smallness of this effect is a combination of the large orbital distance taken to be within the plane wave limit, which decreases the fraction of the pulsar's luminosity that effectively undergoes scattering, and the small value of the maximal gain factor for the superradiant modes, which dominate close to $\theta_S=0, \pi$. When the pulsar is away from the poles, and also for all angles in the counter-rotating case, non-superradiant modes are dominant in the incoming flux, such that the pulsar's luminosity is attenuated.

For comparison, we also include in Fig.~\ref{modulation_em} the results for $\tilde{a}=0.9$ and $\omega_S=0.87\Omega_H$, which again corresponds to the frequency that yields the maximum gain factor for the $l=m=1$ mode for this black hole spin parameter. The results are analogous, but the variation of the pulsar's luminosity is a much smaller effect, with a maximum amplification of only one part in $10^5$. This is a generic trend, i.e.~for smaller black hole spins the effects of superradiance are less pronounced and the pulsar's luminosity will present a smaller angular variation. 


\subsubsection{Gravitational radiation}

As mentioned above, neutron stars are expected to exhibit at least small deviations from a spherical shape, and particularly from axial symmetry. These could be due to their solid crust that supports anisotropic stresses, in conjuction with a possibly liquid interior, magnetic distortions when there is a misalignment of the magnetic and rotation axes and other non-axisymmetric instabilities that have been discussed in the literature (see e.g.~\cite{Bonazzola:1995rb} and references therein). Deviations from axial symmetry will then lead to the emission of gravitational waves due to the non-trivial variation of the neutron star's mass quadrupole moment.  

We will follow here the formalism developed in \cite{Bonazzola:1995rb} to describe a pulsar as a spinning ellipsoid whose principal axis of inertia is misaligned with its rotation axis. The gravitational radiation produced by a source in flat space can be obtained via the well-known quadrupole formula, giving for the metric perturbation components in the transverse-traceless (TT) gauge:
\begin{eqnarray} \label{quadrupole_formula}
h_{ij}={2G/c^4\over |\mathbf{r}-\mathbf{r}_S|}\left[P_i^{\phantom{i} k}P_j^{\phantom{j} l}-{1\over2}P_{ij}P^{kl}\right]\ddot{Q}_{jk}\!\left(t-{|\mathbf{r}-\mathbf{r}_S|\over c}\right),\nonumber\\\!\!
\end{eqnarray}
where in cartesian coordinates $P_{ij}=\delta_{ij}-n_in_j$ is the transverse projection operator, with $n^i$ denoting the unit vector along the direction of the neutron star and $Q_{ij}$ the pulsar's mass quadrupole moment. The latter is defined in terms of the quadrupolar part of the $1/r^3$ term in the large radius expansion of the metric coefficient $g_{00}$ in an asymptotically cartesian and mass centered (ACMC) coordinate system. For a non-relativistic source with a weak gravitational field, this can be expressed in terms of the trace-free part of the moment of inertia tensor $I_{ij}$:
\begin{eqnarray} \label{quadrupole_tensor}
Q_{ij}&=&-I_{ij}+{1\over3}I_k^{\phantom{k} k}\delta_{ij}~, \nonumber\\
 I_{ij}&=&\int d^3x \rho(x_k x^k\delta_{ij}-x_ix_j)~.
\end{eqnarray}
For small deformations of the neutron star, this can be decomposed into a term due to rotation and a term due to deformation. If the pulsar does not precess, which will be the case if most of the star is in a liquid phase as modern dense matter calculations have shown, only the latter term will contribute to the emission of gravity waves and we may therefore discard the former. If we assume that, whatever their origin, deformations of the neutron star induce a preferred direction and rotate with the pulsar, there will be a cartesian coordinate system in the weak-field near zone where the quadrupole moment takes the form:
\begin{eqnarray} \label{quadrupole_tensor_hat}
Q_{\hat{i}\hat{j}}=\left(
\begin{tabular}{c c c}
$-Q_{\hat{z}\hat{z}}/2$ & $0$ & $0$ \\
$0$ & $-Q_{\hat{z}\hat{z}}/2$ & $0$\\
$0$ & $0$ & $Q_{\hat{z}\hat{z}}$
\end{tabular}\right)~.
\end{eqnarray}
To obtain the quadrupole moment in the frame where the pulsar is rotating we simply have to rotate the tensor above. If the star is rotating about the $z$-axis (the black hole's spin axis), which makes a fixed angle $\beta$ with the preferred axis $\mathbf{e}_{\hat{z}}$, this yields:
\begin{eqnarray} \label{quadrupole_tensor_derivative}
\ddot{Q}_{ij}&=&{3\over 2}Q_{\hat{z}\hat{z}} \omega_S^2s_\beta\nonumber\\
&\times&\!\!\!\left(
\begin{tabular}{c c c}
$2s_\beta\cos(2\omega_S t)$ & $2s_\beta\sin(2\omega_S t)$ & $c_\beta \sin(\omega_S t)$ \\
$2s_\beta\sin(2\omega_S t)$ & $-2s_\beta\cos(2\omega_S t)$ & $- c_\beta \sin(\omega_S t)$\\
$ c_\beta \sin(\omega_S t)$ & $- c_\beta \cos(\omega_S t)$ & $0$
\end{tabular}\right)\ ,\nonumber\\
\end{eqnarray}
where $s_\beta\equiv \sin\beta$ and $c_\beta\equiv \cos\beta$. From this we immediately conclude that the pulsar will emit gravitational radiation at frequencies $\omega_S$ and $2\omega_S$, the former being favored if the distortion axis is quasi-aligned with the rotation axis ($\beta\simeq 0, \pi$) and the latter when these axes are perpendicular ($\beta=\pi/2$). Note that for exact alignment no gravitational radiation is produced, since the quadrupole moment is static in this case. It is useful to decompose the quadrupole moment in its two frequency modes and write it in the form:
\begin{eqnarray} \label{quadrupole_tensor_complex}
\ddot{Q}_{ij}&=&{3\over 2}Q_{\hat{z}\hat{z}} \omega_S^2s_\beta \left[-c_\beta e^{-i\omega_S t}T_{ij}^{(1)}+2s_\beta e^{-2i\omega_S t} T_{ij}^{(2)}\right]\nonumber\\
&+& \mathrm{c.c.}
\end{eqnarray} 
where
\begin{eqnarray} \label{quadrupole_basis}
T_{ij}^{(1)}=\left(
\begin{tabular}{c c c}
0 & 0 & $\mp i$\\
0 & 0 & 1\\
$\mp i$ & 1 & 0
\end{tabular}\right)~,\qquad
T_{ij}^{(2)}=\left(
\begin{tabular}{c c c}
1 & $\pm i$ &0\\
$\pm i$ & $-1$ & 0\\
0 & 0 & 0
\end{tabular}\right)\!\!.
\end{eqnarray}
Note that here we have considered both an alignment and an anti-alignment between the pulsar and black hole's spins, corresponding to the upper and lower signs in the above expressions, respectively.

By replacing this into the quadrupole formula (\ref{quadrupole_formula}), we may then compute the NP scalar $\psi_4$ via Eq.~(\ref{NP_scalars_grav}) associated with the gravitational radiation emitted by the spinning neutron star. When the pulsar is far away from the black hole, $L\gg |\mathbf{r}|$, we obtain the plane wave form in Eq.~(\ref{psi4_plane}) with circular polarization factors:
\begin{eqnarray} \label{pol_pulsar_GW}
|h_R^{(1)}| &=& {4\sqrt{\pi} G\epsilon I \omega_S^2 s_\beta c_\beta\over 5 c^4 L}|{}_{-2}Y_{2, \pm1}(\theta_0, \phi_0)|~, \nonumber\\
|h_L^{(1)}| &=& {4\sqrt{\pi} G\epsilon I \omega_S^2 s_\beta c_\beta\over 5 c^4 L}|{}_{-2}Y_{2, \mp1}(\theta_0, \phi_0)|~, \nonumber\\
|h_R^{(2)}| &=& {8\sqrt{\pi} G\epsilon I \omega_S^2 s_\beta^2\over 5 c^4 L}|{}_{-2}Y_{2, \pm2}(\theta_0, \phi_0)|~, \nonumber\\
|h_L^{(2)}| &=& {8\sqrt{\pi} G\epsilon I \omega_S^2 s_\beta^2\over 5 c^4 L}|{}_{-2}Y_{2, \mp 2}(\theta_0, \phi_0)|~, 
\end{eqnarray}
where we have defined the pulsar's ellipticity as:
\begin{eqnarray} \label{ellipticity}
\epsilon\equiv -{3\over 2}{Q_{\hat{z}\hat{z}}\over I}~,
\end{eqnarray}
with $I$ denoting its moment of inertia with respect to the rotation axis. Hence, as for the magnetic dipole emission, we find that the polarization of the gravitational waves emitted by the neutron star depends on its angular position and that they can be chirally polarized both for the single and double frequency emission. The circular polarization components are, in particular, given by $s=-2$ spherical harmonics with $l=2$, which is a direct consequence of the quadrupolar nature of the source. We note also that the azimuthal number characterizing these spin-weighted harmonics is $\pm 1$ ($\pm 2$) for emission at (twice) the pulsar's rotation frequency. 

If instead of taking the limit of large orbital radius we take $r\rightarrow +\infty$ in computing $\psi_4$, we can use Eq.~(\ref{incoming_outgoing_energy_GW}) to determine the total luminosity of the pulsar in gravity waves, yielding for the two frequency channels:
\begin{eqnarray} \label{gw_power_modes}
P_{GW}^{(1)}=c^2_\beta P_{GW}~, \qquad P_{GW}^{(2)}=64s^2_\beta P_{GW}~,
\end{eqnarray}
where
\begin{eqnarray} \label{gw_power}
P_{GW}&=&{2\over5}{G\omega_S^6 I^2\epsilon^2\over c^5}s^2_\beta\nonumber\\
&\simeq & 2\times10^9 s_\beta^2\! \left({1\ \mathrm{ms}\over T_P}\right)^{6}\!\!\left({M_P\over 1.4 M_\odot}\right)^2\!\!\left(R_P\over 10\ \mathrm{km}\right)^4\nonumber\\
&\times &\left({\epsilon\over 10^{-5}}\right)^2 L_\odot~,
\end{eqnarray}
where $M_P$ is the pulsar's mass and we have chosen typical values for neutron star parameters. Comparing with the result given above for the power emitted in magnetic dipole radiation (for comparable inclination factors), we see that gravitational waves can yield a significant contribution to the pulsar's spin down rate for ellipticities $\epsilon\gtrsim 10^{-6}$, which are already being probed by Advanced LIGO \cite{Aasi:2013sia, Abbott:2016tvg}. Conversely, the observed spin-down rates for the known pulsars can then be used to place upper limits on the ellipticity, which for typical milisecond pulsars will be of this order.

Using the above total luminosities in each channel and Eq.~(\ref{incident_flux_GW}), we may then write the incident flux in each gravitational wave multipole in the simple form:
\begin{eqnarray} \label{incoming_mode_energy_pulsar_GW}
{dE_{in}^{(l,m)}\over dt}\! =\!\pi{\lambda^2\over L^2}\left|{}_{-2}Y_{2,\pm(1,2)}(\theta_0,\phi_0)\right|^2\!\left|{}_{-2}Y_{lm}(\theta_0,\phi_0)\right|^2\!P_{GW}^{(1,2)}\nonumber\\
\end{eqnarray}
for the positive frequency modes, while for the negative frequency modes we obtain an analogous result with $\pm \rightarrow \mp$. This is completely analogous to the electromagnetic case, with the difference that in the gravitational case the incoming energy is modulated by $s=-2$ and $l=2$ harmonics. This justifies the generic expression for the incident flux in each pulsar mode first given in \cite{Rosa:2015hoa}:
\begin{eqnarray} \label{incoming_mode_energy_pulsar_generic}
{dE_{in}^{(l,m)}\over dt} =\pi{\lambda^2\over L^2}\left|{}_{-s}Y_{s,\pm m_P}(\theta_0,\phi_0)\right|^2\left|{}_{-s}Y_{lm}(\theta_0,\phi_0)\right|^2P_{s}\nonumber\\
\end{eqnarray}
with $s=1$ for the electromagnetic and $s=2$ for the gravitational wave emission, with $m_P=\omega/\omega_S$ for the emission frequency $\omega$ and $P_s$ the total pulsar luminosity in the corresponding channel.

We may then proceed as for the electromagnetic case to determine the output of the scattering process for gravitational waves, and in particular the relative changes in the pulsar luminosity for different angles $\theta_0$. Let us focus on the double frequency mode, which dominates for $\beta\sim \pi/2$, i.e.~when the pulsar's spin axis and principal axis of inertia are nearly perpendicular.

\begin{figure}[htbp]
\includegraphics[scale=1]{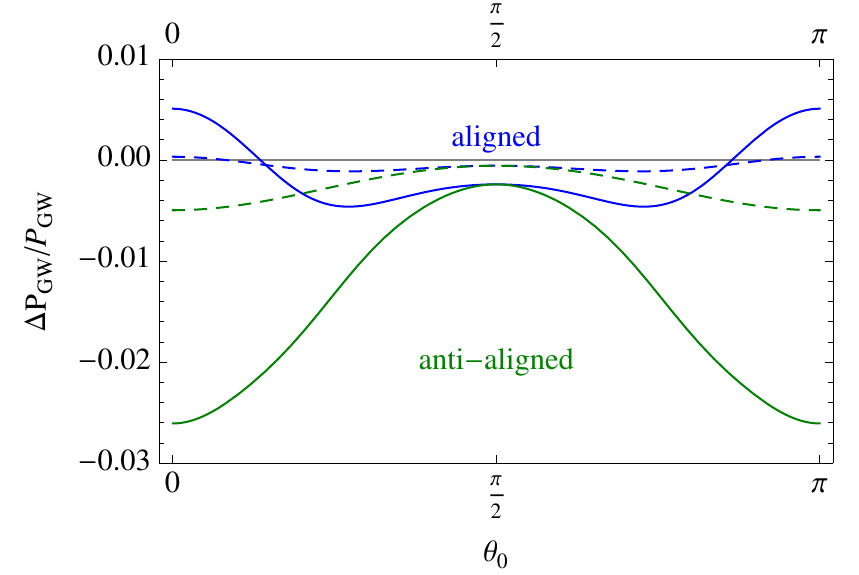}
\caption{Fractional variation of the pulsar's total gravitational wave luminosity due to a near-extremal black hole companion with $\tilde{a}=0.999$, for a pulsar rotational frequency $\omega_S\simeq 0.98\Omega_H$ (solid curves) and for $\tilde{a}=0.9$, $\omega_S\simeq 0.93\Omega_H$ (dashed curves). The blue (green) curves correspond to a co-rotating (counter-rotating) binary system, and in all cases the orbital distance was taken to be $L=10\lambda$.}
\label{modulation_gw}
\end{figure}

In Fig.~\ref{modulation_gw} we show an example with a nearly extremal black hole, $\tilde{a}=0.999$ and a pulsar with angular velocity $\omega_S\simeq 0.98\Omega_H$, such that the $l=m=2$ superradiant mode is maximally amplified. As in the electromagnetic case, we take $L=10\lambda$ to ensure the plane wave approximation's validity. We show the fractional variation of the pulsar's luminosity $\Delta P_{GW}/P_{GW}$ for this case and also for an example with a smaller black hole spin, $\tilde{a}=0.9$, and $\omega_S\simeq 0.93\Omega_H$, again maximizing the $l=m=2$ mode amplification. We note that an accurate computation of this quantity requires considering the multipoles of $l=2-6$, while the contribution of higher multipoles can be safely neglected\footnote{We had neglected the contributions of the $l>3$ modes in \cite{Rosa:2015hoa}, but this was due to an accuracy issue in the numerical calculations that has now been overcome. The $l=4-6$ modes do not, however, change the magnitude of the effect significantly.}.

As for the pulsar's magnetic dipole radiation, the pulsar's gravitational luminosity depends on the incidence angle, and hence on the pulsar's angular position. The effect is very similar in both the electromagnetic and gravitational cases, but with the latter exhibiting significantly larger variations, namely as a consequence of the larger amplification factors of the superradiant modes, which dominate in the co-rotating case when the pulsar is located close to the black hole's spin axis. The minimal angular deviation from the equatorial plane required for an overall amplification of the plane gravitational waves is of $65.6^\circ$ for the near-extremal example, increasing to $75^\circ$ for $\tilde{a}=0.9$. Superradiant modes never dominate in a counter-rotating system (anti-aligned pulsar and black hole spins) and the angular modulation of the pulsar's luminosity is naturally more pronounced for larger black hole spins, potentially reaching percent level differences for a near-extremal black hole and about an order of magnitude less for $\tilde{a}=0.9$.

We note that in both the electromagnetic and gravitational cases one may achieve the maximal amplification effects for mili-second pulsars orbiting stellar mass black holes. For example, with the shortest pulsar period observed of 1.4 ms, we have $\omega_S\simeq 4.5$ kHz, such that $\omega_S\sim \Omega_H$ for a nearly extremal Kerr black hole of about 23$M_\odot$ according to Eq.~(\ref{black_hole_rotation}), which is close to the mass range inferred from the black hole merger events recently observed with Advanced LIGO \cite{Abbott:2016blz}. We thus conclude that potentially observable signals may occur in realistic systems, even though a black hole-neutron star binary has yet to be detected. We note that finding such systems is one of the particular science goals proposed for the Square Kilometer Array observatory \cite{Aharonian:2013av}.


\subsection{Compact binary - supermassive black hole systems}

Binary systems are the canonical textbook example of gravitational wave sources. The radiation is emitted at a frequency of (twice) the orbital frequency of the system, and hence generically at much lower frequencies than the pulsar radiation considered above. On the one hand, this means that this radiation may be within the superradiant regime for stellar mass black holes, but with $\omega_S\ll \Omega_H$ and hence very small amplification factors as we have seen earlier. On the other hand, for a binary system which is itself orbiting a supermassive black hole (SMBH), we may be able to attain $\omega_S\sim \Omega_H$. For example, the SMBH that is believed to reside at the centre of our galaxy, Sagittarius $\mathrm{A}^\star$, has an inferred mass of $4.6\times10^6 M_\odot$, such that $\Omega_H < 0.02$ Hz depending on its still poorly constrained spin. Other SMBH candidates may attain even larger masses, with for instance the one in the M87 galaxy having an inferred mass of $6.6\times10^9M_\odot$, which gives $\Omega_H\lesssim 10^{-5}$ Hz (see e.g.~\cite{Ricarte:2014nca}). This means that binary systems with periods ranging from a few minutes to a few hours may emit gravitational radiation that may be maximally amplified by a supermassive black hole. Such small periods require binary separations of the order of $10^{-3}-10^{-1}$ AU, thus requiring compact binary systems involving e.g. neutron stars and stellar mass black holes. For example, the well-known Hulse-Taylor double neutron star system has a period of 7.75 hours \cite{Hulse:1974eb}.

It is very easy to apply the formalism developed in the previous section to see that the gravitational radiation emitted by a binary system can be amplified by a SMBH companion if the binary and SMBH rotation axes are aligned. Let us consider the simplest (and canonical) case of a system of two stars with equal mass $M$ in a circular orbit of radius $R$ and angular frequency $\omega_S$ (in the slow inspiral phase, well before coalescence). Let us also assume for simplicity that the orbital plane coincides with the SMBH's equatorial plane. In this case, we have for the mass density in a cartesian frame with origin at the binary center of mass:
\begin{eqnarray} \label{binary_mass}
\rho&=& M\delta(z)\left[\delta(x-R\cos(\omega_St))\delta(y-R\sin(\omega_St))\right.\nonumber\\
&+&\left. \delta(x+R\cos(\omega_St))\delta(y+R\sin(\omega_St))\right]~.
\end{eqnarray}
Plugging this into Eq.~(\ref{quadrupole_tensor}), it is easy to obtain:
\begin{eqnarray} \label{quadrupole_tensor_derivative_binary}
\ddot{Q}_{ij}&=&-4MR^2\omega_S^2\left(
\begin{tabular}{c c c}
$2\cos(2\omega_S t)$ & $\sin(2\omega_S t)$ & $0$ \\
$\sin(2\omega_S t)$ & $-\cos(2\omega_S t)$ & $0$\\
$ 0$ & $0$ & $0$
\end{tabular}\right)\ ,\nonumber\\
&=&-2MR^2\omega_S^2 e^{-2i\omega_St} T_{ij}^{(2)} + \mathrm{c.c.}~,\\\nonumber
\end{eqnarray}
where $T_{ij}^{(2)}$ was defined in Eq.~(\ref{quadrupole_basis}). Hence, the quadrupole moment for the compact binary has the same tensor structure as the double frequency mode of the rotating ellipsoid model used for the pulsar example, so that apart from the overall constants one obtains essentially the same result for the circular polarization of the emitted gravity waves, namely that:
\begin{eqnarray} \label{pol_pulsar_GW}
|h_R^{(2)}| &=& {8\sqrt{\pi} GMR^2\omega_S^2 \over 5 c^4 L}|{}_{-2}Y_{2, \pm2}(\theta_0, \phi_0)|~, \nonumber\\
|h_L^{(2)}| &=& {8\sqrt{\pi} GMR^2\omega_S^2 \over 5 c^4 L}|{}_{-2}Y_{2, \mp 2}(\theta_0, \phi_0)|~.
\end{eqnarray}
If we consider an inclined orbit for the compact binary we also obtain a single frequency mode with the same tensor structure as in the pulsar case. Hence, the neutron star and the compact binary lead to essentially the same results apart from the overall amplitude of the waves and the different frequency range, but one will obtain exactly the same variation of the compact binary luminosity as its angular position changes with respect to the SMBH's spin axis. In particular, when the binary co-rotates with the supermassive black hole, superradiant modes will be dominant when the binary is parametrically close to the SMBH's axis, as shown in Fig.~\ref{modulation_gw}.

There is already a significant number of SMBH candidates in AGN observations with very large spin parameters, which would of course be ideal for superradiant amplification of the gravitational radiation emitted by a binary companion \cite{Brenneman:2011wz}. This includes at least three candidates with spins $\geq 0.98$ and masses in the $5\times 10^6M_\odot-10^7M_\odot$ range. Large amplification factors (potentially just below the percent as shown for the analogous pulsar-BH system) could then be obtained if a compact binary with angular frequency $\omega_S\simeq \Omega_H\sim 10^{-2}$ Hz is orbiting such a SMBH at sufficiently small radius and in a sufficiently inclined orbit with respect to the SMBH equatorial plane. Interestingly, such gravitational wave frequencies are within the range of the eLISA mission (see e.g.~\cite{Vitale:2014sla}), so that one may hope that in the near future one could be able to detect the effects of a nearby spinning SMBH on the gravitational radiation emitted by an inspiraling binary.


\subsection{Orbital modulation of the source luminosity}

We have seen above two examples of sources that emit radiation which, in the plane wave limit, can be amplified by a nearby spinning black hole, depending on the source's angular position in the black hole's reference frame. Amplification requires the source to rotate in the same direction as the black hole and to be located away from the equatorial plane, either towards the north or the south pole of the black hole. Orbits in the Kerr space-time can have a complex evolution, with variations both in the radial and angular Boyer-Lindquist coordinates (see e.g.~\cite{Singh:2014nta}). While we plan to perform a more thorough study of superradiance effects for different orbits of the source, here we will focus on the simplest case of Keplerian orbits, which are a good approximation to realistic orbits if the source is sufficiently far away from the black hole. We note that this is, in any case, required by the plane wave approximation considered. 

Let us first consider circular orbits, such that $|\mathbf{r_S}|=L$ is constant. Orbits in the equatorial plane have $\theta_0=\pi/2$, for which the source's radiation is mostly absorbed by the black hole, as we have seen above. Since this angle is constant for these orbits (in the Keplerian limit), the source luminosity will be constant and no interesting effects will arise in this case. We thus need orbits with a significant inclination with respect to the black hole's equatorial plane, such that $|\theta_0-\pi/2|$ exceeds the amplification thresholds computed above. 

To obtain the source's angular position as a function of time for a circular orbit, we may rotate an equatorial orbit $(x,y,z)=\left(L\cos(2\pi t/T),L\sin(2\pi t/T), 0\right)$ by e.g.~an angle $\gamma$ about the $x$ axis to obtain:
\begin{eqnarray} \label{inclined_circ_orbit}
x&=& L\cos(2\pi t/T)~, \nonumber\\
y &=& L\cos\gamma \sin(2\pi t/T)~, \nonumber\\
 z&=& L\sin\gamma \sin(2\pi t/T)~.
\end{eqnarray}
This gives for the source's angular position:
\begin{eqnarray} \label{angle_circ}
\theta_0 (t) &=& \pi - \arccos(z/L)=\nonumber\\
&=&\pi -\arccos \left(\sin\gamma \sin(2\pi t/T)\right)~,
\end{eqnarray}
which we may substitute in Eq.~(\ref{incoming_mode_energy_pulsar_generic}) to compute the incident energy in each multipole as a function of time and, multiplying by the corresponding gain/loss factors, obtain the time variation of the source's luminosity.

\begin{figure}[htbp]
\includegraphics[scale=1]{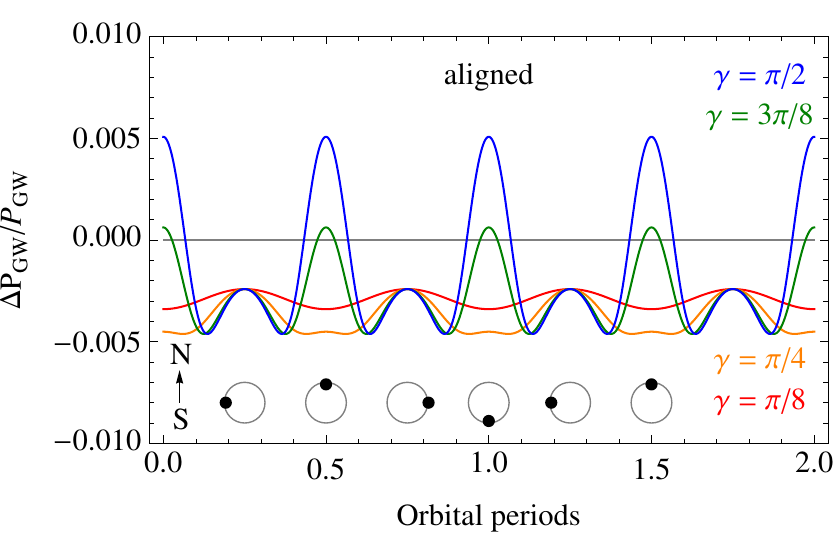}\vspace{0.6cm}
\includegraphics[scale=1]{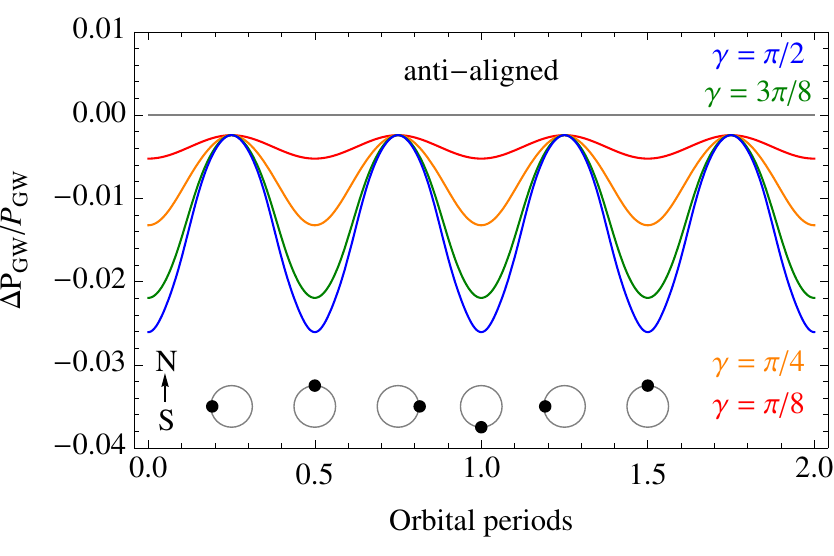}
\caption{Orbital modulation of the source's total gravitational wave luminosity due to a near-extremal black hole companion with $\tilde{a}=0.999$, for $\omega_S\simeq 0.98\Omega_H$ and circular orbits of radius $L=10\lambda$ and different inclination $\gamma$. The top (bottom) plot corresponds to the case where the source and black hole spins are aligned (anti-aligned). The diagrams at the bottom of the figures illustrate the source's orbital position in the case of a polar orbit ($\gamma=\pi/2$).}
\label{orbital_modulation_gw}
\end{figure}

In Fig.~\ref{orbital_modulation_gw} we show the orbital modulation of the gravitational luminosity of a source (either a pulsar or a compact binary) with $\omega_S\simeq 0.98\Omega_H$ orbiting a near-extremal black hole, $\tilde{a}=0.999$, in a circular orbit of radius $L=10\lambda$ and different inclination angles. The top (bottom) plot shows the results for the case where the source and black hole spins are (anti-)aligned.

As one can see in this figure, in the co-rotating case orbits with a small inclination yield only a small variation of the luminosity, since the polar angle $\theta_0$ exhibits smaller orbital variations and never reaches large enough or small enough values for superradiant modes to dominate. One observes only a simple modulation of the pulsar's luminosity at the orbital frequency for the smaller inclination angles. As one increases the inclination, a secondary modulation starts appearing, with peaks at the largest and smallest angular values, where superradiant modes are dominant. For inclinations exceeding $65.6^\circ$ the luminosity is effectively amplified at these peaks, with the largest effect observed for the polar orbit where the source passes right above and below the north and south poles, respectively. A distinctive signature of superradiance is thus a double luminosity modulation for large orbital inclinations in a co-rotating system.

When the source and black hole spins are anti-aligned one always observes a single modulation of the luminosity, and an overall amplification is never attained, since superradiant modes never dominate in this configuration.

Eccentric orbits can also yield distinctive modulation patterns. For an orbit of eccentricity $e$ and semi-major axis $L$, we have $|\mathbf{r}_S|= L/(1+e\cos\phi)$, where $\phi$ denotes the azimuthal angle in the orbital plane. For an inclined orbit we then obtain:
\begin{eqnarray} \label{inclined_eccen_orbit}
x&=& {L\over 1+e\cos\phi}\cos\phi~,\nonumber\\
y &=& {L\over 1+e\cos\phi}\cos\gamma \sin\phi~, \nonumber\\
 z&=& {L\over 1+e\cos\phi}\sin\gamma \sin\phi,
\end{eqnarray}
such that
\begin{eqnarray} \label{angle_circ}
\theta_0 (\phi) &=& \pi - \arccos(z/|\mathbf{r}_S|)=\nonumber\\
&=& \pi -\arccos \left(\sin\gamma \sin\phi\right)~.
\end{eqnarray}

\begin{figure}[t]
\includegraphics[scale=1]{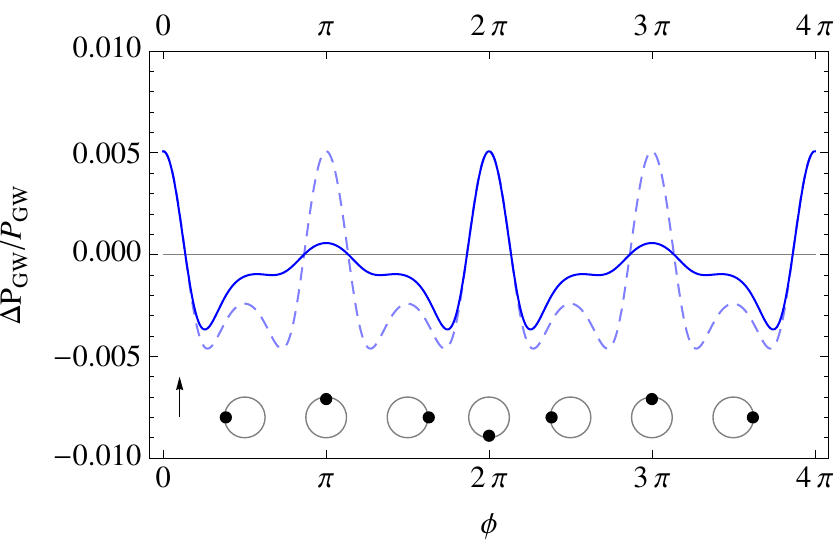}
\caption{Orbital modulation of the source's total gravitational wave luminosity due to a near-extremal black hole companion with $\tilde{a}=0.999$, for $\omega_S\simeq 0.98\Omega_H$ for a polar orbit with eccentricity $e=0.5$ and periastron at $|\mathbf{r}_S|=10\lambda$ coinciding with the passage below the black hole's south pole (solid curve). For comparison we include the results for a circular orbit of radius $L=10\lambda$ (dashed curve). The diagrams at the bottom of the figure illustrate the source's orbital position.}
\label{orbital_modulation_gw_eccentric}
\end{figure}

In Fig.~\ref{orbital_modulation_gw_eccentric} we compare the modulation of the gravitational luminosity of the source considered in Fig.~\ref{orbital_modulation_gw} for a circular and an eccentric polar orbit in the co-rotating case. We consider a configuration where the periastron occurs when the source is above the black hole north pole. Naturally, the variation of the luminosity is larger when the source is closer to the black hole, and for the configuration chosen this enhances one of the peaks corresponding to superradiant amplification, attenuating the effect in the remainder of the orbit.  If, however, the periastron does not coincide with the source's passage close to one of the poles, the effects of superradiance will be attenuated. Nevertheless, if a configuration such as the one shown in Fig.~\ref{orbital_modulation_gw_eccentric} can be found in Nature, one may hope to more easily identify the effects of superradiance.

\vspace{0.5cm}
In the examples shown above, integrating the luminosity variation over an orbital period yields a negative result, implying that overall the wave deposits more energy and spin into the black hole than it extracts from the latter. While we cannot discard the possibility that in more complicated orbits there is a net energy and spin extraction, it is unlikely that this results in any measurable effect. Since only a small fraction of the source's energy flux effectively interacts with the black hole, the time required for extracting a significant amount of mass and spin from the latter will necessarily exceed the source's lifetime. In particular, a pulsar will lose its rotational energy much faster than it spins down a black hole companion and a compact binary will also merge before having any significant effect on a supermassive black hole. We thus expect the modulation of the source's luminosity to be the leading observational effect of the presence of a spinning Kerr black hole.

Another issue to note is the fact that the orbital motion of the quadrupolar source, either the pulsar or the compact binary in the examples studied above, will itself lead to the generation of gravitational waves at (twice) the orbital frequency. The adiabatic approximation that we have used is justified for orbital frequencies $\Omega_{orb}\ll \omega_S$, but this only means that we may take the source as approximately fixed in analyzing the gravity waves produced by the intrinsic rotation of the source. It is thus important to estimate the effect of the orbital component. Let us consider, for example, the case of a pulsar-black hole binary. In this case, the typical strain amplitude of the gravity waves produced by the orbital motion in the inspiral regime can be written as:
\begin{eqnarray} \label{h_orbit}
h_0^{orb} = {4G/c^4\over r}{GM_P M_{BH}\over L}~,
\end{eqnarray}
for a circular orbit of radius $L$ and angular frequency $\Omega_{orb}= \sqrt{G(M_{BH}+M_P)/L^3}$.  Now, the typical strain amplitude corresponding to the gravity waves emitted by the pulsar due to rotation at $\omega=2\omega_S$ is:
\begin{eqnarray} \label{h_pulsar}
h_0^{spin} = {4G/c^4\over r}{2\over 5} \epsilon M_{P}R_P^2\omega_S^2~,
\end{eqnarray}
so that we obtain:
\begin{eqnarray} \label{h_ratio}
{h_0^{spin}\over h_0^{orb}} \simeq 2 \epsilon \left({R_{NS}\over 10\ \mathrm{km}}\right)^2\!\!\left( 23M_\odot\over M_{BH}\right)\!\!\left({T_P\over 1\ \mathrm{ms}}\right)\!\!\left({L\over 10\lambda}\right).
\end{eqnarray}
Hence, the amplitude of the gravitational waves due to the orbital motion may largely exceed the rotational component for small ellipticities. This could pose a challenge for the observation of the rotational component and its modulation due to scattering off the black hole companion, which encodes the effects of superradiance. However, we note that, for the parameters chosen above, $\Omega_{orb}\sim $ few Hz, while $\omega_S\sim $ few kHz, so that there is a clear frequency separation that may allow one to isolate the smaller rotation component in a Fourier decomposition using filtering techniques. The fact that the amplitude of the latter should be modulated at the orbital frequency, as our analysis of the total luminosity suggests, does not constitute a problem, since in Fourier space the signal modulation only introduces spectral components at frequencies $2(\omega_S\pm \Omega_{orb})\gg \Omega_{orb}$, so that e.g.~applying a high-pass filter should eliminate the orbital component while preserving the modulated rotational signal.


\section{Conclusion}

In this work we have shown that there are realistic astrophysical systems where superradiant amplification of electromagnetic and gravitational radiation by a spinning black hole may occur, with potential astrophysical signatures. We started by analyzing the generic conditions under which plane electromagnetic or gravitational waves can be amplified by a Kerr black hole, corresponding to the limit where the radiation source is far away from the latter. We concluded that plane waves must satisfy three necessary conditions: (i) have a sufficiently low frequency $\omega \lesssim  \Omega_H$, (ii) the incidence angle must be parametrically close to the black hole's rotation axis and (iii) the right/left polarizations of the wave must be dominant at small/large incidence angles. 

We have then considered potential astrophysical sources of low-frequency radiation, namely a spinning neutron star or pulsar, which emits both magnetic dipole and quadrupolar gravitational radiation, and a compact binary system, where only the latter is emitted. We have shown that in both cases the third condition for amplification can be satisfied when the source co-rotates with the black hole. Superradiance can, in these cases, lead to a characteristic double modulation of the source's total luminosity as it orbits the black hole, which is more pronounced for larger black hole spins and orbital inclinations, and the closer the source's angular frequency is to the horizon's angular velocity (which maximizes the amplification factors of the leading superradiant modes). Our results for gravitational radiation are completely analogous for the pulsar and the compact binary, and in fact the same luminosity modulation should occur for any spinning quadrupolar source, and in the electromagnetic case our results should apply to any spinning dipole, whether magnetic or electric in nature.

Since we have considered the plane wave limit of the incident radiation, also performing the corresponding mode decomposition in the flat space limit, our results are limited to orbital distances that are large compared to the wavelength of the radiation and the horizon radius. Naturally, the effects of superradiance are larger the closer the source is to the black hole, which motivates a further study of this process beyond the plane wave approximation. In the co-rotating case, at least, we may expect that the relative fraction of superradiant modes in the incident wave becomes larger as the source approaches the black hole, thus enhancing the effects of superradiance, although this requires a more detailed analysis that we hope to perform in the future.

The pulsar-binary black hole system is quite appealing from the observational perspective, since one could hope for observational effects of superradiance in both the electromagnetic and the gravitational channels. In fact, amplification factors are largest in both channels when the pulsar's angular frequency $\omega_S\simeq \Omega_H$ (noting that gravitational radiation is emitted also at frequency $2\omega_S\sim 2\Omega_H$). However, the low frequency magnetic dipole radiation emitted by a spinning neutron star poses considerable challenges. Firstly, the pulsar's magnetosphere is thought to be filled with an ionized plasma, since the strong electric field associated with the magnetic dipole radiation pulls charged particles from the neutron star's surface, thus creating a plasma in its vicinity. These particles are then accelerated along the magnetic field lines (the `pulsar wind') and produce the beamed radio emission that we observe from the Earth. 

It is widely believed that most of the energy of the magnetic dipole radiation, which cannot propagate below the plasma frequency, is efficiently converted into accelerating these particles and generating higher frequency radiation, in which case it may not even reach the black hole and scatter in the way we have analyzed. There are nevertheless examples of pulsars where this plasma may be absent at least for limited periods. For example, the pulsar B1931+24 is only radio-active for periods of 5-10 days, remaining ÒquietÓ for periods of 25-35 days. Measurements of the spin-down rate in these two phases are consistent with a main energy loss mechanism through pulsar wind in the active phase and through magnetic dipole radiation in the quiet phase \cite{Gurevich}. It would then be possible for the magnetic dipole radiation to scatter of a black hole companion in these quiet phases.

A further challenge for electromagnetic signals is that the scattered waves will necessarily interact with different ionized plasmas as they travel towards the Earth.  The Earth's atmosphere is itself opaque to electromagnetic radiation with frequencies below $\sim$10 MHz and the Moon may have an ionosphere with plasma frequency in the few hundred kHz range. This would imply that to detect radio waves with only a few kHz one would need to go into the outer Solar System, with the Voyager spacecrafts having detected radiation down to $2-3$ kHz. The interstellar and intergalactic medium are also filled with ionized plasma, with plasma frequencies of a few kHz and a few Hz (at small redshifts), respectively. Although this could allow for the propagation of magnetic dipole radiation from a mili-second pulsar, low frequency waves are significantly affected by free-free absorption and other opacity sources and are mostly converted into heat \cite{Lacki:2010jr}. 

This suggests that low-frequency electromagnetic radiation may only be observed indirectly through the way it heats up intervening gas clouds along the line of sight. In this case, the luminosity modulation due to the pulsar's orbit around the spinning black hole may potentially be indirectly inferred through the consequent modulation of the temperature of a known gas cloud. The known periodicity of this effect could in principle make it easier to distinguish it from other astrophysical sources but this is, of course, a challenging task.

Gravitational waves yield a much more promising channel for observing the effects of black hole superradiance. Not only are amplification factors larger than in the electromagnetic case but they also hardly interact with astrophysical plasmas or other sources that could potentially mask the effects of superradiance. Of course this makes them more difficult to detect, but the recent results of the LIGO and VIRGO collaborations have shown that gravitational wave astronomy is a reality \cite{Abbott:2016blz}, and better sensitivities will be reached in the near future.  

The analysis performed in this work is not yet fully realistic in the sense that spinning neutron stars are more complex objects than precessing mass quadrupoles (or magnetic dipoles as discussed above). The emission of gravity waves by pulsars can have a variety of sources other than rotation (see e.g.~\cite{Kokkotas:1999bd, Lasky:2015uia}) and one must investigate whether these other components can also undergo superradiant scattering or if the component due to rotation can be observationally disentangled from other low-frequency modes. Similarly, the waves emitted by an inspiraling compact binary also change in frequency and shape as the system approaches coalescence, although within the range of validity of the plane wave approximation our results should yield a good approximation.

Furthermore, we have focused on the effects of superradiance on the total luminosity of a source orbiting a Kerr black hole, and observationally one can only measure the luminosity along a particular line-of-sight. In addition, in the case of gravitational waves, current detector technology is sensitive to strain rather than luminosity. Since the power in gravitational waves is quadratic in the strain amplitude, one may expect a larger relative modulation of the latter and potentially signals that are easier to observe, but this requires a careful analysis that, along with the above-mentioned line-of-sight effects, we plan to perform in the future. We will also extend our analysis to more general spin and orbital configurations.

Nevertheless, the results in this work point towards the existence of realistic systems where superradiance may induce non-trivial effects in the radiation emitted by realistic sources orbiting both stellar mass and supermassive black holes, with a magnitude potentially large enough to be observed in a not too distant future. Of course one must hope that Nature provides us with systems with the right properties - a highly-spinning black hole orbited by a co-rotating pulsar/compact binary with an angular frequency $\omega\sim \Omega_H$, in a sufficiently inclined orbit. There is considerable evidence for the existence of these individual components, and in principle they may exist in binary systems as well. 
Hence, observing black hole superradiance may become a reality in the future and certainly constitute a very important test of general relativity and of the nature of compact astrophysical bodies.

\begin{acknowledgments}

I would like to thank Vitor Cardoso and Carlos Herdeiro for their useful comments and suggestions. This work was supported by the FCT Grant No.~SFRH/BPD/85969/2012 and partially by the H2020-MSCA-RISE-2015 Grant No.~StronGrHEP-690904, and by the CIDMA Project No.~UID/MAT/04106/2013.

\end{acknowledgments}


\appendix

\section{Spin-weighted spherical harmonics}

Spin-$s$ spherical harmonics ${}_sY_{lm}(\theta,\phi)$, $l\geq |s|$, can be obtained from the spherical harmonic functions $Y_{lm}(\theta,\phi)\equiv{}_0Y_{lm}(\theta,\phi)$ by applying the spin raising and lowering differential operators \cite{Goldberg:1966uu}:
\begin{eqnarray} \label{spin_weighted_harmonics}
\eth {}_sY_{lm}&=&\sqrt{l(l+1)-s(s+1)}{}_{s+1}Y_{lm}~,\nonumber\\
\bar\eth {}_sY_{lm}&=&-\sqrt{l(l+1)-s(s-1)}{}_{s-1}Y_{lm}~,
\end{eqnarray}
where
\begin{eqnarray} \label{spin_operators}
\eth\eta =-\left(\partial_\theta+{i\over\sin\theta} \partial_\phi-s\cot\theta\right)\eta~,\nonumber\\
\bar\eth\eta =-\left(\partial_\theta-{i\over\sin\theta} \partial_\phi+s\cot\theta\right)\eta~,
\end{eqnarray}
with $\eta$ denoting an arbitrary spin-weight $s$ function, and the normalization was chosen such that:
\begin{eqnarray} \label{spin_harmonics_int}
\int_0^{2\pi}\!\! d\phi \int_0^\pi d\theta\sin\theta {}_sY_{lm}^* {}_sY_{l'm'}^*= \delta_{ll'}\delta_{mm'}~.
\end{eqnarray}
The spin-$s$ spherical harmonics satisfy ${}_sY_{lm}^*=(-1)^{m+s} {}_{-s}Y_{lm}$. We list below the lowest spherical harmonics with spin weight $s=-1$:
\begin{eqnarray} \label{spin_harmonics_1}
{}_{-1}Y_{10}&=& -\sqrt{3\over 8\pi} \sin\theta~,\nonumber\\
 {}_{-1}Y_{1\pm1}&=& -\sqrt{3\over 16\pi}( \cos\theta\pm1)e^{\pm i\phi}~,\nonumber\\
 {}_{-1}Y_{20}&=& -\sqrt{15\over 8\pi} \sin\theta\cos\theta~,\nonumber\\
  {}_{-1}Y_{2\pm1}&=& -\sqrt{5\over 16\pi} (\cos2\theta \pm \cos\theta)e^{\pm i\phi}~,\nonumber\\
    {}_{-1}Y_{2\pm2}&=& \sqrt{5\over 16\pi} \sin\theta(1\pm \cos\theta)e^{\pm 2i\phi}~,
\end{eqnarray}
and $s=-2$:
\begin{eqnarray} \label{spin_harmonics_2}
{}_{-2}Y_{20}&=& \sqrt{15\over 32\pi} \sin^2\theta~,\nonumber\\
 {}_{-2}Y_{2\pm1}&=& \sqrt{5\over 16\pi}\sin\theta( \cos\theta\pm1)e^{\pm i\phi}~,\nonumber\\
 {}_{-2}Y_{2\pm 2}&=& -\sqrt{5\over 64\pi} (1\pm\cos\theta)^2e^{\pm 2i\phi}~,\nonumber\\
 {}_{-2}Y_{30}&=& \sqrt{105\over 32\pi} \sin^2\theta\cos\theta~,\nonumber\\
{}_{-2}Y_{3\pm1}&=&\sqrt{35\over 128\pi} \sin\theta\left(\cos\theta\pm 1\right)\left(3\cos\theta\mp 1\right)e^{\pm i\phi}~,\nonumber\\
{}_{-2}Y_{3\pm2}&=&-\sqrt{21\over 128\pi} \sin\theta\left(\cos\theta\pm 1\right)^2e^{\pm 3i\phi}~.
\end{eqnarray}
%


\section{Multipolar decomposition}

As shown above, the NP scalars for electromagnetic and gravitational plane waves $\phi_2$ ($s=1$) and $\psi_4$ ($s=2$) in flat space can be written in the generic form:
\begin{eqnarray} \label{NP_1}
\Upsilon_s &=& A_s(\gamma_R e^{-i\omega t +i \mathbf{k}\cdot\mathbf{r}}+ \gamma_Le^{i\omega t -i \mathbf{k}\cdot\mathbf{r}})\nonumber\\
&\times& \sum_{m_s} {}_{-s}Y_{sm_s}^*(\mathbf{\hat{k}}) {}_{-s}Y_{sm_s}(\mathbf{\hat{r}})~,
\end{eqnarray}
with $\mathbf{\hat{k}}=(\theta_0,\phi_0)$ and $\mathbf{\hat{r}}=(\theta,\phi)$. We may then use the scalar plane wave multipolar decomposition in Eq.~(\ref{scalar_harmonics})  to obtain, for the positive frequency part:
\begin{eqnarray} \label{NP_2}
\Upsilon^+_s &=& 4\pi A_s\gamma_R e^{-i\omega t }\!\sum_{l, m_l, m_s} i^l j_l(kr) Y^*_{lm_l}(\mathbf{\hat{k}}){}_{-s}Y^*_{sm_s}(\mathbf{\hat{k}})\nonumber\\
&\times & Y_{lm_l}(\mathbf{\hat{r}}){}_{-s}Y_{sm_s}(\mathbf{\hat{r}})~.
\end{eqnarray}
Products of spin-weighted spherical harmonics admit a Clebsch-Gordan decomposition of the form:
\begin{eqnarray} \label{CG_1}
{}_{s_1}Y_{l_1m_1}\, {}_{s_2}Y_{l_2m_2}= \sum_{jm} A_{jm} {}_sY_{jm}~,
\end{eqnarray}
with $|l_1-l_2|\leq j \leq l_1+l_2$ and Clebsch-Gordan coefficients:
\begin{eqnarray} \label{CG_2}
A_{jm}&=&\sqrt{(2l_1+1)(2l_2+1)\over 4\pi (2j+1)} \langle l_1,s_1; l_2, s_2 | j, s\rangle \nonumber\\
&\times & \langle l_1, m_1; l_2, m_2 | j, m\rangle \delta_{m, m_1+m_2} \delta_{s, s_1+s_2}~.
\end{eqnarray}
We may then use this to write the NP scalars as:
\begin{eqnarray} \label{NP_3}
\Upsilon^+_s &=&\!\!  (2s+1)A_s\gamma_R e^{-i\omega t }\!\!\sum_{l, m_l, m_s}\!\sum_{j,m,j',m'}\!\!{i^l (2l+1)j_l(kr)\over \sqrt{2j+1(2j'+1)}}\nonumber\\
&\times& \langle l, 0; s,-s | j, -s\rangle \langle l, m_l; s,m_s | j, m\rangle\nonumber\\
&\times& \langle l, 0; s,-s | j', -s\rangle^* \langle l, m_l; s,m_s | j', m'\rangle^*\nonumber\\
&\times & {}_{-s}Y_{j'm'}^*(\mathbf{\hat{k}}) {}_{-s}Y_{jm}(\mathbf{\hat{r}})~.
\end{eqnarray}
The sum over $m_l$ and $m_s$ can be performed using the Clebsch-Gordan identity:
\begin{eqnarray} \label{CG_3}
\sum_{m_1, m_2} \langle l_1, m_1; l_2, m_2| j, m\rangle \langle  j', m'| l_1, m_1; l_2, m_2\rangle = \delta_{jj'} \delta_{mm'} \nonumber\\
\end{eqnarray}
which then gives:
\begin{eqnarray} \label{NP_4}
\Upsilon^+_s &=&\!\!  (2s+1)A_s\gamma_R e^{-i\omega t }\sum_{l,j,m}i^lj_l(kr){2l+1\over 2j+1}\nonumber\\
&\times& |\langle l, 0; s,-s | j, -s\rangle|^2 {}_{-s}Y_{jm}^*(\mathbf{\hat{k}}) {}_{-s}Y_{jm}(\mathbf{\hat{r}})~.
\end{eqnarray}
To obtain the large distance form of the NP scalars in Eqs.~(\ref{phi2_large}) and (\ref{psi4_large}) we can use that:
\begin{eqnarray} \label{CG_4}
 |\langle l, 0; s,-s | j, -s\rangle|^2 &=& {(2j+1)(2s)!\over(l+s-j)!(l+s+j+1)!}\nonumber\\
 &\times & {(j+s)! (j+l-s)!\over (j-s)! (j-l+s)!}
\end{eqnarray}
to perform the sum over $l$ of the spherical Bessel functions in the limit $kr\gg 1$, which can be easily done e.g.~using Mathematica. An analogous procedure can then be used to obtain the asymptotic expansion for the negative frequency modes.




\end{document}